\documentclass[11pt,a4paper]{article}
\usepackage{jcappub}
\usepackage{bm}
\usepackage{bbm}
\usepackage{amsmath}
\usepackage{etoolbox}
\usepackage{float}
\usepackage{tikz}
\usepackage{braket}
\usepackage{ulem}

\newcommand{\bp}{{\bf p}}

\newcommand{\bk}{{\bf k}}
\newcommand{\bl}{{\bf l}}

\newcommand{\btl}{{\tilde{\bf l}}}
\newcommand{\btk}{{\tilde{\bf k}}}

\defcitealias{Friman:1979ecl}{FM}

\begin{document}
\hfill CERN-TH-2024-092
\subheader{\date{\today}}

\title{Neutrino emission in cold neutron stars: Bremsstrahlung and
modified urca rates reexamined
}

\author[a]{Salvatore Bottaro}
\author[b]{Andrea Caputo}
\author[c]{Damiano F. G. Fiorillo}

\makeatletter
\@date
\makeatother

\affiliation[a]{School of Physics and Astronomy, Tel-Aviv University, \\ Tel-Aviv 69978, Israel}
\affiliation[b]{Department of Theoretical Physics, CERN, Esplanade des Particules 1, P.O. Box 1211, Geneva 23, Switzerland}
\affiliation[c]{Niels Bohr International Academy, Niels Bohr Institute, University of Copenhagen, 2100 Copenhagen, Denmark}

\emailAdd{salvatoreb@tauex.tau.ac.il}
\emailAdd{andrea.caputo@cern.ch}
\emailAdd{damiano.fiorillo@nbi.ku.dk}

\abstract{
Neutrino emission in cold neutron stars is dominated by the modified urca (murca) process and nucleon-nucleon bremsstrahlung. The standard emission rates were provided by Friman and Maxwell in 1979, effectively based on a chiral Lagrangian framework with pion and rho meson exchange, supplemented by Landau parameters to describe short-range interactions. We reevaluate these rates within
the same framework, correcting several errors and removing
unnecessary simplifications, notably the triangular approximation -- where the Fermi momenta of protons and leptons negligible compared to that of neutrons -- in MURCA, and quantify their importance.
The impact of rho meson exchange, previously argued to cancel with interference effects, is actually quite relevant. Altogether, the cooling rates are reduced by as much as a factor 2. We provide comprehensive analytical formulas encompassing all contributions, designed for straightforward numerical implementation. Our results are particularly relevant for studies of physics beyond the standard model, where the emission of new particles – such as axions – is typically computed within the same framework we adopt here.
}

\maketitle

\section{Introduction}

Neutron stars (NS) are fascinating objects. It was 1932 when -- just one month \textit{before} the discovery of the neutron -- Landau conjectured the existence of cold dense stars in a conversation with Bohr and Rosenfeld. These ``scary stars" (``unheimliche Sterne") would have small radii and enormous densities, as also suggested independently in 1934 by Baade and Zwicky. A few years later, in 1939, Oppenheimer and Volkoff described the first model of neutron stars, made of a free neutron gas, but it was only in 1967 that pulsars were observed and shortly after Gold proposed the now accepted view that they are fast rotating neutron stars. Typical neutron star masses vary in the range $M_{\rm NS} \sim 1-2 \, M_\odot$ with a radius of the order of $R \sim 10 \, \rm km$, therefore with very large densities, \textit{beyond} nuclear, $\rho_0 \sim 2.5 \times 10^{14} \, \rm g/cm^3$ or $n_0 \sim 0.15 \, \rm fm^{-3}$.

These extreme stars form when the degenerate iron core of a massive star at the end of its life evolution becomes unstable and collapses, leading to a type II supernova explosion. The first stages of the life of the newborn are decisively determined by neutrino emission, which dominates the cooling of the neutron star. At the beginning, the inner temperature is very high, around $T \sim 10^{10}-10^{11} \, \rm K = 1-10 \, \rm MeV$, but it quickly drops to $T \sim 10^9 \, \rm K \sim 100 \, \rm keV$ after 1-10 years. After $\sim 10^4-10^5$ years, the temperature drops to $T \sim 2 \times 10^8 \, \rm K$, and surface photon emission finally takes over in the cooling process. Clearly, neutrino cooling processes in nuclear environments are of paramount importance for the study of these fascinating stars. The first detailed computations of neutrino emission rates in neutron stars trace back to 50 years ago~\cite{Dicus:1972yr, Bahcall:1965zzb, Flowers:1975ac}. Usually, the direct decay of a neutron $n\to p+ e^-+\overline{\nu}_e$ is inhibited by the strong degeneracy, so the most important processes to consider are neutrino-antineutrino bremsstrahlung
\begin{eqnarray}\label{Eq:BremProcess}
    n + n \rightarrow n + n + \nu_{\ell} + \bar{\nu}_{\ell}, \\
    n + p \rightarrow n + p + \nu_{\ell} + \bar{\nu}_{\ell}, \\
    p + p \rightarrow p + p + \nu_{\ell} + \bar{\nu}_{\ell},
\end{eqnarray}
where $\ell =e, \mu, \tau$ indicates the neutrino flavor, and modified URCA (MURCA) processes, where a charged lepton is also emitted, both in the neutron branch 
\begin{eqnarray}\label{Eq:MURCAProcess}
    n + n \rightarrow n + p + \ell + \bar{\nu}_{\ell}, \quad
    n + p + \ell \rightarrow n + n + \nu_{\ell}\ ,
\end{eqnarray}
and the proton branch
\begin{eqnarray}\label{Eq:MURCAProcessPB}
    p + n \rightarrow p + p + \ell + \bar{\nu}_{\ell}, \quad
    p + p + \ell \rightarrow p + n + \nu_{\ell}.
\end{eqnarray}
The standard reference for the emission rates associated with these processes is the seminal paper written by Friman and Maxwell in 1979~\cite{Friman:1979ecl} (hereafter~\citetalias{Friman:1979ecl}), in which the authors provided a detailed description of all the computations within the one-pion exchange (OPE) framework, but quantifying also the effect of short-distance interactions beyond OPE. The topic has been reconsidered later on, for example by Yakovlev and Levenfish~\cite{Yakovlev1995A&A} with a focus on reduction factors due to possible proton superfluidity, and by Maxwell himself with a focus on the potential presence of hyperons~\cite{Maxwell1987ApJ}. Even without superfluidity, the use of in-vacuum interactions has been questioned, as the medium polarization might significantly renormalize the properties and interactions of nucleons~\cite{Schwenk:2003pj,Schwenk:2003bc, Shternin:2018dcn, vanDalen:2003zw, Bacca:2008yr, DehghanNiri:2016cqm, Blaschke:1995va,Sedrakian:2024uma}. 
Nevertheless, standard reviews~\cite{Yakovlev:2000jp} and fast public numerical codes like \href{http://www.astroscu.unam.mx/neutrones/NSCool/}{NSCool}, still use the results from the original work~\citetalias{Friman:1979ecl}, which represents hitherto the most complete and detailed reference for the topic. The ramifications of these results go well beyond pure astrophysics, as NS cooling is a sensitive probe of emission of particles beyond the Standard Model (BSM)~\cite{Buschmann:2021juv}, where it is usually treated using NSCool, in turn relying on~\citetalias{Friman:1979ecl}.

Given the potential impact of neutrino cooling in NS across multiple fields, especially in recent times, we believe a detailed understanding of its rate is needed. Here we reconsider the historical work of~\citetalias{Friman:1979ecl}, and we show that several corrections arising from the purely particle side of the calculation can alter the cooling rates by more than a factor~2. We do not aim for a state-of-the-art description of the nuclear physics (in particular, we do not tackle in-medium effects, which are known to be relevant in NS interiors~\cite{Schwenk:2003pj,Schwenk:2003bc, Shternin:2018dcn, vanDalen:2003zw, Bacca:2008yr, DehghanNiri:2016cqm, Blaschke:1995va,Sedrakian:2024uma, Blaschke:2013vma}), but stick to the framework adopted in~\citetalias{Friman:1979ecl}. Our results correct for multiple small inconsistencies in the original treatment, at the same time relieving some of the approximations of the original paper; specifically:

\begin{itemize}
    \item we model the long-range interaction as an OPE+$\rho$-meson exchange. The latter was argued in~\citetalias{Friman:1979ecl} to compensate with the exchange contribution, at least for the $nn$ bremsstrahlung, but we do not seem to recover this result. The impact of the $\rho$-meson exchange recently entered also the particle physics community, where it was argued to significantly affect the emission rate of axions~\cite{Carenza:2019pxu}, although here it was incorrectly accounted for in the non-tensor channels contributing to $np$ bremsstrahlung. We take the occasion to rectify the impact of this contribution; 
    \item for the first time, we go beyond the triangular approximation that the Fermi momentum of protons and electrons are negligible; one can already see that this is usually smaller than the neutron Fermi momentum only by about a factor $3$ or less, so it is worth considering more carefully what is the impact of this approximation. 
\end{itemize}

When all is put together, neutrino emission rates, compared to the results in~\citetalias{Friman:1979ecl}, differ by a factor~$2$ or more in some cases. Even though our results may not be the final answer, since we still stick with the nuclear framework of~\citetalias{Friman:1979ecl}, it is a direct update of the most commonly used approaches to describe NS cooling, especially in the context of bounds on BSM physics. Thus, we believe that our results could be an updated go-to recipe in this context.


This work is organized as follows. We first introduce our framework, describing the adopted nucleon-nucleon potential. Then, we pass to compute the emission rates for both bremsstrahlung and MURCA. For each of them we provide contact with previous literature and we highlight the impact of different effects on the final answer. We finally provide a compact expression also for the neutrino absorption rate and then conclude. 


\section{Nuclear physics framework}

The emission of weakly interacting particles from dense nuclear matter is notoriously challenging to describe. The intrinsic many-body nature of the system does not allow, in principle, to discuss properties of individual nucleons. However, one can usually adopt the Landau paradigm to describe such a system of strongly interacting fermions by means of non-interacting quasi-particles, representing collective degrees of freedom of the system which behave nearly as individual particles. We will adopt this viewpoint in what follows.

The main challenge in using Landau's theory of Fermi liquids is that the properties of individual quasi-particles -- mass, coupling to neutrinos -- become phenomenological, to be determined from comparison with experiment. Here, in order to stick to a well-defined framework, we will make a set of simplifying assumptions, mostly driven by the comparison with the often-used work by Friman and Maxwell~\citetalias{Friman:1979ecl}. In particular, we will describe the nuclear matter in terms of non-interacting, non-relativistic nucleons. The dispersion relation of quasi-particles close to their Fermi surface is determined by their Landau effective masses. For the results shown in the text, we will always use bare values for the masses to provide a comparison with the results of~\citetalias{Friman:1979ecl} which does not include potential differences from this additional source of uncertainty. All emissivities scale with a fixed power of the nucleon mass, so introducing a definite prescription for the effective masses can always be done in a simple way.
In asymmetric nuclear matter, the proton effective mass may be lower than the neutron one, an effect we do not consider in this work, aiming at a precision of the order of $10\%$. This choice is especially helpful in comparing our results with the classic FM ones.

The rate of neutron star cooling is now determined by emission of neutrinos from nucleons. On-shell emission processes from an individual nucleon could only happen by direct neutron decay $n\to p+e^- +\overline{\nu}_e$ and inverse beta decay $p+e^-\to n+\nu_e$, but both processes are strongly suppressed by the degeneracy of dense nuclear matter. Hence, neutrino emission happens mostly from off-shell nucleons interacting with each other. This implies the need for a detailed discussion of the quasi-particle interaction in the Fermi liquid of nucleons. Unfortunately, this is a topic clouded in unavoidable complexity and uncertainties. 

The forward scattering amplitude in various interaction channels might be related, using Landau's theory of a Fermi liquid, to specific thermodynamical coefficients (e.g. compressibility, spin susceptibility, ...) which can in principle be measured in heavy nuclei, although one should stress that such measurements always refer to symmetric nuclear matter, whereas neutron stars are obviously neutron-dominated. Here we follow FM, which used the Fermi liquid parameters extrapolated from Refs.~\cite{Sjoberg:1976tq,Anastasio:1977et}. For low momentum exchange, such that $kr_0\ll 1$, where $k$ is the typical momentum transfer and $r_0$ is the typical nucleon radius, the amplitude is isotropic, and the Landau parameters are directly representative of the scattering amplitude. We stress that these parameters directly map to the scattering amplitude, not the scattering potential; hence, they allow us to circumvent the need for a perturbative treatment of the interaction potential.

The Landau parameters are by definition unable to capture the effects of the tensor interaction driven primarily by pion exchange, which vanish for vanishing momentum transfer. However, tensor interactions are the dominant contribution for most neutrino-emission processes, and in fact are the only contribution for $nn$ bremsstrahlung. Here, we assume that the long-range tensor interaction can be described as in vacuum by the pion exchange, with a reduction at shorter scales which we model as an exchange of a $\rho$ meson. In this sense, we follow again~\citetalias{Friman:1979ecl}, except that we extract the coupling of the $\rho$ meson from the Bonn potential~\cite{Machleidt:1987hj}. It must also be noted that assuming the in-vacuum interaction is not quite adequate, since the medium polarization could significantly renormalize the tensor interaction~\cite{Schwenk:2003pj,Schwenk:2003bc, Shternin:2018dcn, vanDalen:2003zw, Bacca:2008yr, DehghanNiri:2016cqm, Blaschke:1995va}; in Ref.~\cite{Schwenk:2003pj}, for example, the resulting $nn$ bremsstrahlung emission was shown to be renormalized by up to a factor~$2$\footnote{However, in Eq.~3 of Ref.~\cite{Schwenk:2003pj} a symmetry factor of $2$, rather than $4$, is reported for $nn$ bremsstrahlung. Presumably a clear estimate of the impact of these medium corrections requires some dedicated analysis, also in view of the corrections we point out in this paper, which we do not attempt here.}. Our choice of sticking to the simpler framework of~\citetalias{Friman:1979ecl} allows us to perform a one-to-one comparison with their results to show that significant differences arise already at this level, while a detailed treatment of the medium polarization would certainly be warranted in future works.

Hence, our final framework to describe nucleon-nucleon interaction includes short-range interactions inferred from the Landau parameters, and longer range contributions from OPE+$\rho$-meson exchange. We find that the former have negligible impact on neutrino-neutrino emission, a feature already identified in~\citetalias{Friman:1979ecl}. On the other hand, as we will see, $\rho$ exchange does significantly affect the emission. Overall, the effective non-relativistic potential that we use for nucleon-nucleon interaction is
\begin{equation}\label{Eq:GenPotential}
V(\mathbf{k}) = f + f' \boldsymbol{\tau}_1 \cdot \boldsymbol{\tau}_2 + g \boldsymbol{\sigma}_1 \cdot \boldsymbol{\sigma}_2 + g'_k \boldsymbol{\tau}_1 \cdot \boldsymbol{\tau}_2 \, \boldsymbol{\sigma}_1 \cdot \boldsymbol{\sigma}_2  + h'_k  (\boldsymbol{\sigma}_1 \cdot \boldsymbol{\hat{\bk}}) (\boldsymbol{\sigma}_2 \cdot \boldsymbol{\hat{\bk}}) \boldsymbol{\tau}_1 \cdot \boldsymbol{\tau}_2,
\end{equation}
 where ${f, f', g}$ are the constant Landau parameters for the relevant channel and $\bk\equiv k \hat{\bk}$ is the exchanged momentum in the scattering. The spin-spin interaction $g'_k$ receives both a constant contribution from the associated Landau parameter and a momentum-dependent contribution arising from the exchange of the $\rho$
\begin{equation}\label{eq:gk}
    g'_k \equiv g' - C_\rho \, \frac{f_\pi^2}{m_\pi^2} \frac{k^2}{k^2+m_\rho^2}.
\end{equation}
For the meson parameter here we follow the table at pag.~37 for the Bonn model~\cite{Machleidt:1987hj} and take $m_\rho = 769 \, \rm MeV$ for the mass of the $\rho$ meson, and $C_\rho = 1.4$ for its coupling strength, while $f_\pi \simeq 1$ and should \textit{not} be confused with the pion decay constant. Notice that scattering data have been also used to compute the emission of neutrinos and exotic particles in Core Collapse Supernovae~\cite{Hanhart:2000er, Hanhart:2000ae, Rrapaj:2015wgs}.

For the Landau parameters instead we adopt the following parametrisation~\cite{Sjoberg:1976tq, Anastasio:1977et, Speth:1980kw} $\{f, f', g, G\} = \frac{\pi^2}{2 \, m_{\rm N} \, p_{\rm F}(n)} \{F_0, F_0', G_0, G'_0\}$, where $F_0' = 0.7, G_0 = G_0' = 1.1$ and where $p_{\rm F}$ is the neutron Fermi momentum. As already noted by \citetalias{Friman:1979ecl}, and as we confirm in our results, the Landau parameter $F_0$ drops in all the relevant rates. Finally, the last operator in Eq.~\ref{Eq:GenPotential} reads
\begin{equation}\label{eq:hk}
    h'_k \equiv - \frac{f_\pi^2}{m_\pi^2} \Big(\frac{k^2}{k^2 +m_\pi^2} - C_\rho \frac{k^2}{k^2 +m_\rho^2}\Big),
\end{equation}
where we stress the sign difference between the pion and $\rho$ contributions. As a matter of fact, Ref.~\cite{Ericson:1988wr} included $\rho$-meson exchange only in the tensor coupling $h'_k$, but this was justified by the choice of only determining $nn$ bremsstrahlung, where tensor interaction is the only contribution. 

This was subsequently followed by Ref.~\cite{Carenza:2019pxu} which adopted the simple rule 
\begin{equation}
    \frac{k^2}{k^2 +m_\pi^2}\to \frac{k^2}{k^2 +m_\pi^2} - C_\rho \frac{k^2}{k^2 +m_\rho^2},
\end{equation}
which however is wrong if applied to the spin-spin channel. Hence, the corresponding results for $np$ scattering in Ref.~\cite{Carenza:2019pxu} -- which usually is the dominant contribution to axion emission in supernovae -- do not consistently incorporate the $\rho$-meson exchange reduction. Since here we consider not only $np$ bremsstrahlung but also MURCA, we account for the consistent prescription.

We find it useful -- especially for the particle physicist reader -- to stress that the contributions due to OPE+$\rho$ exchange can be derived from the following Lagrangian

\begin{equation}
\mathcal{L} = - i g_\pi \bar{N}\gamma_5 \tau^a N \pi^a- i \, g_\rho \bar{N}\gamma_\mu  \tau^a N \rho^{a,\mu}- i \, \frac{f_\rho}{4\, m_{\rm N}}\bar{N}\sigma_{\mu\nu} \tau^a  N \Big(\partial^\mu\rho^{a,\nu} - \partial^\nu\rho^{a,\mu}\Big) ,
\end{equation}

where $\tau^a$ are the usual Pauli matrices, $N = \begin{pmatrix} \chi_p \\ \chi_n \end{pmatrix}$, with $\chi_{\rm p,n}$ being the spinor fields associated with the proton and the neutron, while $\pi^a = \Big(\frac{\pi^+ + \, \pi^-}{\sqrt{2}}, \frac{i \, (\pi^+ - \, \pi^-)}{\sqrt{2}}, \pi^0\Big)$ are the pion fields. Taking the non relativistic limit one can derive the following relations between the coupling constant in this Lagrangian and the parameters for the potential in Eq.~\ref{Eq:GenPotential} 
\begin{equation}
    g_\pi = \frac{2 \, m_{\rm N} f_\pi}{m_\pi}, \, \, C_\rho = \frac{(g_\rho + f_\rho)^2 m_\pi^2}{4\, m_{\rm N}^2} =g_\rho^2\frac{(1 + r)^2 m_\pi^2}{4\, m_{\rm N}^2 },
\end{equation}

where we introduced the tensor to vector ratio of the $\rho$-meson, $r \equiv f_\rho/g_\rho$. Following Ref.~\cite{Machleidt:1987hj} we fix $r = 6.1$ and $g_\rho = \sqrt{4\pi \times  0.41} \simeq 2.3$. The inclusion of the $\rho$ meson primarily serves the purpose of reducing the OPE potential which otherwise at large momentum transfer would saturate to a constant value and largely overpredict the interaction energy. 

Finally, for completeness, we also report the piece of the non-relativistic Lagrangian describing the electroweak interactions; for the charged-current interactions this is 
\begin{equation}\label{Eq:LagrangianEW}
\begin{split}
    \mathcal{L}=\, \,&  \frac{G}{\sqrt{2}}\left(\chi_p^\dagger(\delta^\mu_0-g_A \sigma^i\delta^\mu_i)\chi_n \bar{\ell}\gamma_\mu(1-\gamma_5)\nu_{\ell}+\mathrm{h.c.}\right),
\end{split}
\end{equation}

where $g_A \simeq 1.27$ is the axial vector constant, $G = G_{\rm F} \cos\theta_C$ with $G_{\rm F}$ being the weak Fermi coupling constant and $\theta_C$ the Cabibbo angle, $\chi_{n,p}$ are two-components Pauli spinors while $\ell$ and $\nu_{\ell}$ are standard Dirac spinors. For the neutral-current interactions with neutrinos, we use 
\begin{equation}
    \mathcal{L}= \frac{G_{\rm F}}{2\sqrt{2}}\left(\chi_p^\dagger(c_v\delta^\mu_0-g_A \sigma^i\delta^\mu_i)\chi_p -\chi_n^\dagger(\delta^\mu_0-g_A \sigma^i\delta^\mu_i)\chi_n \right)\overline{\nu}_{\ell}\gamma_\mu (1-\gamma_5)\nu_{\ell}+\mathrm{h.c.},
\end{equation}
with $c_v=1-4\sin^2\theta_W$ and $\theta_W$ is Weinberg's angle.

In concluding this section, we wish to emphasize that the numerical results presented herein utilize the Lagrangian couplings in their bare form, akin to the treatment of nucleon masses. It is essential to note, however, that these couplings are expected to get in-medium modifications. For example, axial coupling $g_A$ is expected to experience quenching at finite densities~\cite{Carter:2001kw}. Our results can be rescaled accordingly.

\section{Neutrino emissivity}

With the nucleon interaction potential at hand, we can proceed to compute the neutrino emissivity for both bremsstrahlung and MURCA. The general form for \textit{single-flavor} neutrino emissivity (we will later account for the proper flavor multiplicities) reads
\begin{equation}\label{Eq:Rate}
Q_\nu = \mathcal{S} \int \, \frac{d^3p_{\nu,1}}{(2\pi)^3 2\, \omega_1} \int \prod_{i=1}^4 \frac{d^3p_i}{(2\pi)^3} \int \frac{d^3 p_{\ell}}{(2\pi)^3 2\, \omega_{\rm l}} \omega_\nu (2\pi)^4 \delta(P_1 + P_2 - P_3 - P_4 - P_{\ell} - Q_\nu) \mathcal{F} \sum_{\rm spins} |\mathcal{M}|^2,
\end{equation}
where $\mathcal{S}$ is a symmetry factor for identical particles, equal to $1/4$ for neutron-neutron and proton-proton bremsstrahlung, $1/2$ for MURCA and 1 for neutron-proton bremsstrahlung. In Eq.~\ref{Eq:Rate}, capital letters $P=(E, \bp)$ denote the 4-momentum of the corresponding particle. In particular, $\bp_i$ and $E_i$ are the nucleon momenta and energies respectively, $\omega_{\ell}$ is the energy of the second lepton emitted (either charged leptons for MURCA or neutral leptons for bremsstrahlung ), and we also defined the phase space factor $\mathcal{F}$, which reads $\mathcal{F}= f_1 f_2 (1-f_3)(1 - f_4)$ for bremsstrahlung and $ \mathcal{F}=f_1 f_2 (1-f_3)(1 - f_4)(1-f_{\ell})$ for MURCA, where $f_i$ are the fermion distribution functions with the appropriate chemical potentials. Finally, $\omega_\nu$ is the total energy emitted into neutrinos (therefore $\omega_\nu = \omega_1 + \omega_\ell$ for bremsstrahlung and $\omega_\nu = \omega_1$ for MURCA). In all that follows, we assume that neutrinos freely escape and we can neglect their distribution functions in the Boltzmann equation, an extremely good approximation already a few seconds after the formation of a neutron star. Moreover, we will work under the assumption of strong degeneracy for all nucleons, as well as for muons and electrons, so that scattering processes only involve fermions close to the Fermi surface with momentum $p_{\rm F,i}$ and width $\sim T/p_{\rm F,i} \ll 1$. For the nucleons, we do \textit{not} include the usual factor $2m_{\rm N}$ in the denominator appearing in a relativistic treatment, which would cancel out with a corresponding factor in the normalization of the relativistic spinors in the matrix element. Thus, we simply consider the nucleon wavefunctions normalized according to the condition $\overline{N}N=1$, more appropriate for non-relativistic calculations. Finally, we do \textit{not} include a factor $2$ for the spin of the particles in the integration over the phase space, which means that in our squared amplitude calculations we always have to sum, not average, over the spins. 

\subsection{Bremsstrahlung emission}

There are two types of processes to consider in this case: one with two identical nucleons (either protons or neutrons) in the initial and final states, and one with a neutron and a proton scattering off each other. We treat the two cases separately as both the amplitudes squared and the phase spaces differ. 

\subsubsection{Neutron-neutron (proton-proton) bremsstrahlung }

The squared amplitude for identical nucleons bremsstrahlung is
\begin{equation*}
|\mathcal{M}|^2 = 64 g_A^2 G_{\rm F}^2 \frac{\omega_1 \omega_2}{(\omega_1+\omega_2)^2} 
\left[ h'^2_l + h'^2_k + h'_kh'_l \left( 1 - 3 \, (\hat{\bk}\cdot \hat{\bl})^2 \right) \right].
\end{equation*}
Here we have already averaged over the directions of the outgoing leptons; since their momenta are much smaller than the nucleon momenta, they are emitted essentially isotropically with uncorrelated directions.
Only the tensor interaction contributes to the squared matrix element, as one can easily understand by noting that neutrino emission couples to the total spin of the nucleon pair, and all the other interactions conserve the total spin. In a more compact form we can write
\begin{equation}
|\mathcal{M}|^2=64g_A^2G_{\rm F}^2\frac{\omega_1\omega_2}{(\omega_1+\omega_2)^2}|m|^2,
\end{equation}
where we introduced the reduced squared amplitude $|m|^2$, defined as 
\begin{eqnarray}
|m|^2=\left(\frac{f_\pi}{m_\pi}\right)^4\Big(\Big[\frac{l^2}{(l^2+m_\pi^2)}-C_\rho \frac{l^2}{(l^2+m_\rho^2)}\Big]^2+\Big[\frac{k^2}{(k^2+m_\pi^2)}-C_\rho \frac{k^2}{(k^2+m_\rho^2)}\Big]^2  \\ + \Big[\frac{l^2}{(l^2+m_\pi^2)}-C_\rho \frac{l^2}{(k^2+m_\rho^2)}\Big]\Big[\frac{k^2}{(k^2+m_\pi^2)}-C_\rho \frac{k^2}{(k^2+m_\rho^2)}\Big] \Big),\nonumber
\end{eqnarray}
where $\bk\equiv \bp_1-\bp_3$, $\bl\equiv \bp_2-\bp_3$, with $\bp_i$ being the spatial momenta of the involved nucleons. In writing the squared amplitude, we used the fact that $\hat{\bk}\cdot\hat{\bl}$ vanishes for nucleons exactly on the Fermi surface, and thus is suppressed for very small $T/\mu_N$.

The integrals over the Fermi distributions are easily evaluated in the limit $T\to 0$, using the result that
\begin{equation}
    \int_{-\infty}^{+\infty}\prod_{i=1}^4 dx_i \frac{\delta(x_1+x_2-x_3-x_4-\xi)}{(e^{x_1}+1)(e^{x_2}+1)(e^{-x_3}+1)(e^{-x_4}+1)}=\frac{1}{1-e^{-\xi}}\frac{2\pi^2\xi}{3}\left(1+\frac{\xi^2}{4\pi^2}\right);
\end{equation}
it proves convenient to reinstate the integral over the nucleon momenta forcing them to be on the Fermi surface. Thus, we can rewrite the emissivity as
\begin{equation}
    Q_\nu=\frac{G_{\rm F}^2 g_A^2 m_{\rm N}^4}{96\pi^{12}}\int \omega_1^2 d\omega_1 \omega_2^2 d\omega_2 \frac{T^2(\omega_\nu^2+4\pi^2T^2)}{1-e^{-\omega_\nu/T}}\prod_{i=1}^4 d^3 p_i\delta(p_i^2-p_{F,i}^2) \delta^{(3)}(\bp_1+\bp_2-\bp_3-\bp_4) |m|^2,
\end{equation}
where we already included the symmetry factor $\mathcal{S}=1/4$. The integrals over the nucleons phase space are strongly constrained by the delta functions. Since the matrix element $|m|^2$ depends only on $|\bk|=|\bp_3-\bp_1|$ and $|\bl|=|\bp_4-\bp_1|$, it is most convenient to reparameterize the phase space integration in terms of these differences. After performing all of the integrals except those on $k=|\bk|$ and $l=|\bl|$, we are left with
\begin{equation}\label{Eq:QBremNN}
\begin{split}
Q_\nu^{\rm NN}=&\frac{g_A^2G_{\rm F}^2m_{\rm N}^4}{48 \pi^{10}}\underbrace{\int_{0}^\infty d\omega_1 d\omega_2 \omega_1^2 \omega_2^2\frac{4\pi^2T^2+(\omega_1+\omega_2)^2}{e^{(\omega_1+\omega_2)/T}-1}}_{\mathcal{I}}\times \underbrace{\int_0^{2p_{\rm F}}dk\int_0^{\sqrt{4p_{\rm F}^2-k^2}}dl\frac{|m|^2}{\sqrt{4p_{\rm F}^2-k^2-l^2}}}_J\\
\equiv &\frac{g_A^2G_{\rm F}^2m_{\rm N}^4}{48 \pi^{10}} \mathcal{I} J,
\end{split}
\end{equation}
where $J\equiv 2J_1+J_2$ and
\begin{equation}
\begin{split}
 &J_1= \frac{ f_\pi^4}{m_\pi^4}\pi p_{\rm F}\left[\phi(\alpha)+\phi(\beta)C_\rho^2-2C_\rho \Phi(\alpha,\beta)\right],\\
    & J_2=\frac{f_\pi^4 }{m_\pi^4}\pi p_{\rm F}\left[\Psi(\alpha,\alpha)-2 C_\rho \Psi(\alpha,\beta)+C_\rho^2 \Psi(\beta,\beta)\right].
\end{split}
\end{equation}

Here $p_{\rm F}$ is the Fermi momentum of either neutrons or protons, $\alpha=m_\pi/2p_{\rm F}$, $\beta=m_\rho/2p_{\rm F}$, and
\begin{equation}
\label{eq:phi1}
    \phi(\alpha)=1+\frac{\alpha^2}{2(1+\alpha^2)}-\frac{3}{2}\alpha\;\mathrm{arctan}\left(\frac{1}{\alpha}\right),
\end{equation}
\begin{equation}
\label{eq:phi2}
    \Phi(\alpha,\beta)=1-\frac{\alpha^3\;\mathrm{arctan}\left(\frac{1}{\alpha}\right)-\beta^3\;\mathrm{arctan}\left(\frac{1}{\beta}\right)}{\alpha^2-\beta^2},
\end{equation}
\begin{equation}
   \Psi(\alpha,\beta)=1-\alpha\;\mathrm{arctan}\left(\frac{1}{\alpha}\right)-\beta\;\mathrm{arctan}\left(\frac{1}{\beta}\right) \nonumber +\frac{\alpha \beta}{\sqrt{1+\alpha^2 +\beta^2}} \mathrm{arctan}\left(\frac{ \sqrt{1+\alpha^2+\beta^2}}{\alpha\beta}\right).
\end{equation}

The energy integral $\mathcal{I}$ in Eq.~\ref{Eq:QBremNN} reads instead
\begin{eqnarray}
\mathcal{I}&=& \int_0^{\infty}d\omega_1 d\omega_2 \omega_1^2 \omega_2^2\frac{4\pi^2T^2+\omega^2}{e^{\omega/T}-1}= \nonumber\\ &=& \frac{1}{32} \int_0^\infty d\omega \int^\omega_{-\omega} d\delta \, \frac{(\omega + \delta)^2(\omega - \delta)^2}{e^{\omega/T}-1}(\omega^2 + 4 \pi^2 T^2) = \frac{164 \pi^8 T^8}{4725},
\end{eqnarray}
and therefore
\begin{equation}\label{Eq:FinalQBremNN}
Q_\nu^{\rm NN} = \frac{41\, G_{\rm F}^2 m_{\rm N}^4 \,  T^8}{56700 \pi^2} \, J.
\end{equation}

\begin{figure}[htp!]
    \centering
    \includegraphics[scale = 0.5]{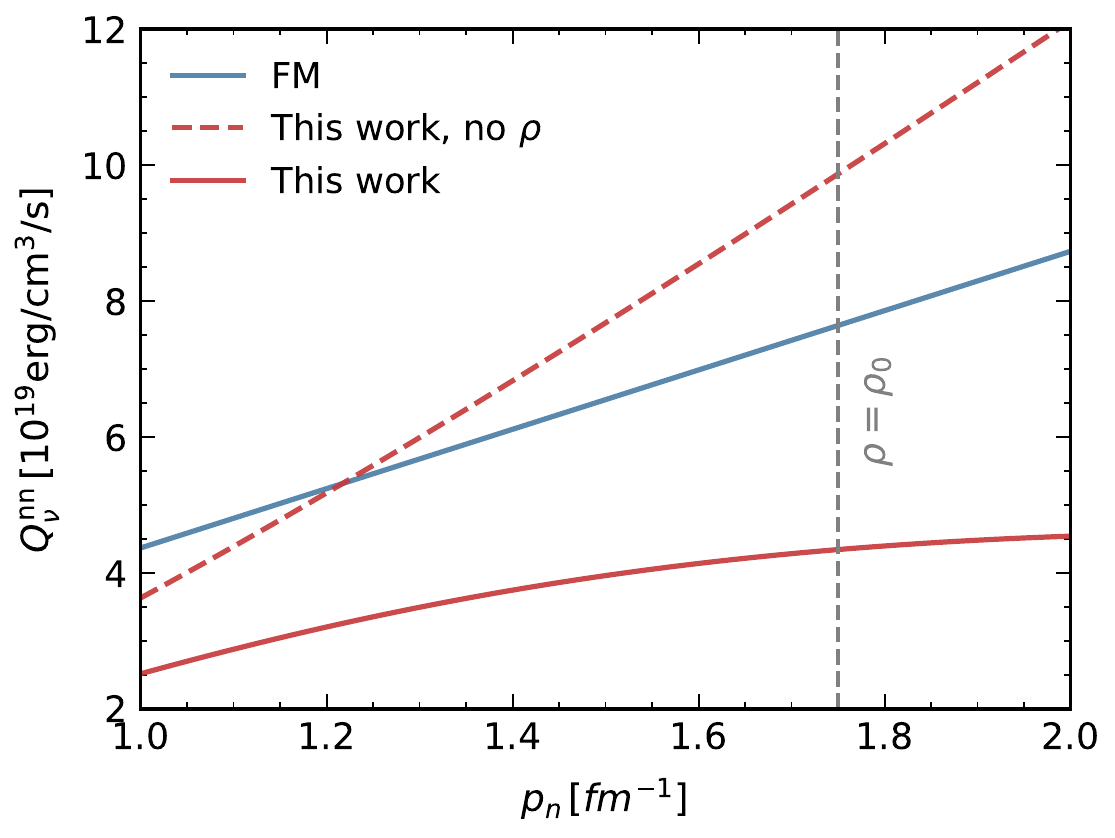}
    \caption{Comparison between the neutron-neutron bremsstrahlung obtained in this work (red curve), and the emissivity computed by \citetalias{Friman:1979ecl} (blue curve) and used in public codes such as NSCool. Our result for nuclear densities (marked by the vertical gray dashed line) and above is considerably smaller. For all curves we have fixed the temperature to $T = 10^9 \, \rm K$, $N_\nu = 3$, the neutron mass to be the bare one and we also included the extra suppression factor ${\tt beta_{nn} = 0.56}$ introduced in Ref.~\cite{Friman:1979ecl}.}
    \label{fig:bremNN_error}
\end{figure}

Eq.~\ref{Eq:FinalQBremNN} includes all diagrams and the exchange of the $\rho$ meson. 
In Fig.~\ref{fig:bremNN_error}, we show our complete result (red curve) and directly compare it with the final numerical results Eq.~65a in \citetalias{Friman:1979ecl}, which is the standard expression used also in public numerical code \href{http://www.astroscu.unam.mx/neutrones/NSCool/}{NSCool}. This latter -- in its public version -- has the following rate for nucleon-nucleon bremsstrahlung in units of $\rm erg/cm^3/s$
\begin{equation}\label{Eq:NSCoolNNBrem}
{\tt qbrem_{nn}=n_\nu \cdot 7.4d19 \cdot mstn(i)^4 \cdot (kfn(i)/1.68d0) \cdot alpha_{nn} \cdot beta_{nn} * (t/1.d9)^8},
\end{equation}
where $\tt t$ is the temperature, normalised here to $10^9 \, \rm K$, $\tt n_\nu$ is the number of neutrinos flavors, ${\tt alpha_{nn}=0.59, beta_{nn} = 0.56}$ are kept constant in the code at their nuclear density values (although they should be a function of density), ${\tt kfn(i)}$ is the nucleon Fermi momentum in units of $\rm fm^{-1}$,and ${\tt mstn(i)}$ represent the nucleon mass in units of the bare one. This expression coincides in fact with Eq.~65a of ~\citetalias{Friman:1979ecl}. In the comparison plot we fix the number of emitted neutrino flavors to 3, so Eq.~\ref{Eq:FinalQBremNN} has been multiplied by $3$; correspondingly, we show the result of Eq.~52 of~\citetalias{Friman:1979ecl} multiplied by~3, or equivalently their Eq.~65a multiplied by $3/2$ since there they accounted for two-flavor emission. For both expressions we fix the neutron mass to be its \textit{in-vacuum} value; in order to implement any in-medium prescription for the effective mass, $\tilde{m}_N = f(p_{\rm F}) m_{\rm N}$, it will be sufficient to scale our results by $f(p_{\rm F})^4$. We also multiply Eq.~\ref{Eq:FinalQBremNN} by the factor ${\tt beta_{nn} = 0.56}$. This is an extra suppression factor introduced by ~\citetalias{Friman:1979ecl} in their OPE potential to capture ``short-range correlations induced by the hard core of the NN interaction" (see their Eq.~17 and Tab.~I).

~\citetalias{Friman:1979ecl} stated that the result in their Eq.~52 was a good representation of the final neutrino emissivity because of a compensation between two missing effects: the ``exchange diagrams" contribution and the $\rho$-exchange suppression. However, we do not observe this compensation and our result for nuclear densities, \textit{i.e} $k_f \sim 1.7 \, \rm fm^{-1} \sim 335 \, \rm MeV$ (indicated by a gray dashed vertical line in Fig.~\ref{fig:bremNN_error}), and above is considerably smaller. Given the simplicity of the master formula \ref{Eq:FinalQBremNN}, it can be easily implemented when doing numerical comparison within the framework of OPE+$\rho$-meson exchange.

\subsubsection{Neutron-proton bremsstrahlung}

\begin{figure}[H]
    \centering
    \begin{minipage}{0.9\textwidth}
        \centering
        \begin{tikzpicture}[scale=1, transform shape]
            \node at (0,0) {(1) $n$};
            \node at (0,-2) {(2) $p$};
            \node at (4,0) {(3) $n$};
            \node at (4,-2) {(4) $p$};
            \draw[thick] (0.5,0) -- (3.5,0);
            \draw[thick] (0.5,-2) -- (3.5,-2);
            \draw[thick, dashed] (2,0) -- (2,-2);
            \node at (2.8,-1) {$\pi^0_k, \rho_k^0$};
            \draw[thick] (1.5,0) -- ++(-0.7,0.5) node [left] {$\bar{\nu}_2$};
            \draw[thick] (1.5,0) -- ++(-0.5,0.7) node [above] {$\nu_1$};
            \node at (2,-2.5) {(a)};
        \end{tikzpicture}
        \hspace{1cm}
        \begin{tikzpicture}[scale=1, transform shape]
            \node at (0,0) {(1) $n$};
            \node at (0,-2) {(2) $p$};
            \node at (4,0) {(3) $n$};
            \node at (4,-2) {(4) $p$};
            \draw[thick] (0.5,0) -- (3.5,0);
            \draw[thick] (0.5,-2) -- (3.5,-2);
            \draw[thick, dashed] (2,0) -- (2,-2);
            \node at (2.8,-1) {$\pi^0_k, \rho_k^0$};
            \draw[thick] (2.5,0) -- ++(0.7,0.5) node [right] {$\nu_1$};
            \draw[thick] (2.5,0) -- ++(0.5,0.7) node [above ] {$\bar{\nu}_2$};
            \node at (2,-2.5) {(b)};
        \end{tikzpicture}
        \hspace{1cm}
        \begin{tikzpicture}[scale=1, transform shape]
            \node at (0,0) {(1) $n$};
            \node at (0,-2) {(2) $p$};
            \node at (4,0) {(3) $n$};
            \node at (4,-2) {(4) $p$};
            \draw[thick] (0.5,0) -- (3.5,0);
            \draw[thick] (0.5,-2) -- (3.5,-2);
            \draw[thick, dashed] (2,0) -- (2,-2);
            \node at (2.8,-1) {$\pi^0_k, \rho_k^0$};
            \draw[thick] (1.5,-2) -- ++(-0.7,-0.5) node [below right=2.5pt] {$\nu_1$};
            \draw[thick] (1.5,-2) -- ++(-0.5,-0.7) node [left=-2.5pt] {$\bar{\nu}_2$};
            \node at (2,-2.5) {(c)};
        \end{tikzpicture}
        \hspace{1cm}
        \begin{tikzpicture}[scale=1, transform shape]
            \node at (0,0) {(1) $n$};
            \node at (0,-2) {(2) $p$};
            \node at (4,0) {(3) $n$};
            \node at (4,-2) {(4) $p$};
            \draw[thick] (0.5,0) -- (3.5,0);
            \draw[thick] (0.5,-2) -- (3.5,-2);
            \draw[thick, dashed] (2,0) -- (2,-2);
            \node at (2.8,-1) {$\pi^0_k, \rho_k^0$};
            \draw[thick] (2.7,-2) -- ++(0.7,-0.5) node [right] {$\nu_1$};
            \draw[thick] (2.7,-2) -- ++(0.5,-0.7) node [below left=-2pt] {$\bar{\nu}_2$};
            \node at (2,-2.5) {(d)};
        \end{tikzpicture}
    \end{minipage}
    
    \begin{minipage}{0.9\textwidth}
        \centering
        \begin{tikzpicture}[scale=1, transform shape]
            \node at (0,0) {(1) $n$};
            \node at (0,-2) {(2) $p$};
            \node at (4,-2) {(3) $n$};
            \node at (4,0) {(4) $p$};
            \draw[thick] (0.5,0) -- (3.5,0);
            \draw[thick] (0.5,-2) -- (3.5,-2);
            \draw[thick, dashed] (2,0) -- (2,-2);
            \node at (2.8,-1) {$\pi^+_l, \rho_l^+$};
            \draw[thick] (1.5,0) -- ++(-0.7,0.5) node [left] {$\bar{\nu}_2$};
            \draw[thick] (1.5,0) -- ++(-0.5,0.7) node [above] {$\nu_1$};
            \node at (1.5,-2.5) {(e)};
        \end{tikzpicture}
        \hspace{1cm}
        \begin{tikzpicture}[scale=1, transform shape]
            \node at (0,0) {(1) $n$};
            \node at (0,-2) {(2) $p$};
            \node at (4,-2) {(3) $n$};
            \node at (4,0) {(4) $p$};
            \draw[thick] (0.5,0) -- (3.5,0);
            \draw[thick] (0.5,-2) -- (3.5,-2);
            \draw[thick, dashed] (2,0) -- (2,-2);
            \node at (2.8,-1) {$\pi^+_l, \rho_l^+$};
            \draw[thick] (2.5,0) -- ++(0.7,0.5) node [right] {$\nu_1$};
            \draw[thick] (2.5,0) -- ++(0.5,0.7) node [above ] {$\bar{\nu}_2$};
            \node at (1.5,-2.5) {(f)};
        \end{tikzpicture}
        \hspace{1cm}
        \begin{tikzpicture}[scale=1, transform shape]
            \node at (0,0) {(1) $n$};
            \node at (0,-2) {(2) $p$};
            \node at (4,-2) {(3) $n$};
            \node at (4,0) {(4) $p$};
            \draw[thick] (0.5,0) -- (3.5,0);
            \draw[thick] (0.5,-2) -- (3.5,-2);
            \draw[thick, dashed] (2,0) -- (2,-2);
            \node at (2.8,-1) {$\pi^+_l, \rho_l^+$};
            \draw[thick] (1.5,-2) -- ++(-0.7,-0.5) node [below right=2.5pt] {$\nu_1$};
            \draw[thick] (1.5,-2) -- ++(-0.5,-0.7) node [left=-2.5pt] {$\bar{\nu}_2$};
            \node at (1.55,-2.5) {(g)};
        \end{tikzpicture}
        \hspace{1cm}
        \begin{tikzpicture}[scale=1, transform shape]
            \node at (0,0) {(1) $n$};
            \node at (0,-2) {(2) $p$};
            \node at (4,-2) {(3) $n$};
            \node at (4,0) {(4) $p$};
            \draw[thick] (0.5,0) -- (3.5,0);
            \draw[thick] (0.5,-2) -- (3.5,-2);
            \draw[thick, dashed] (2,0) -- (2,-2);
            \node at (2.8,-1) {$\pi^+_l, \rho_l^+$};
            \draw[thick] (2.7,-2) -- ++(0.7,-0.5) node [right] {$\nu_1$};
            \draw[thick] (2.7,-2) -- ++(0.5,-0.7) node [below left=-2pt] {$\bar{\nu}_2$};
            \node at (1.5,-2.5) {(h)};
        \end{tikzpicture}
    \end{minipage}

    \caption{Feynman diagram for neutron-proton bremsstrahlung. Diagrams a-d constitute the t-channel, while e-h represent the u-channel. }
    \label{DiagramNPBrem}
\end{figure}
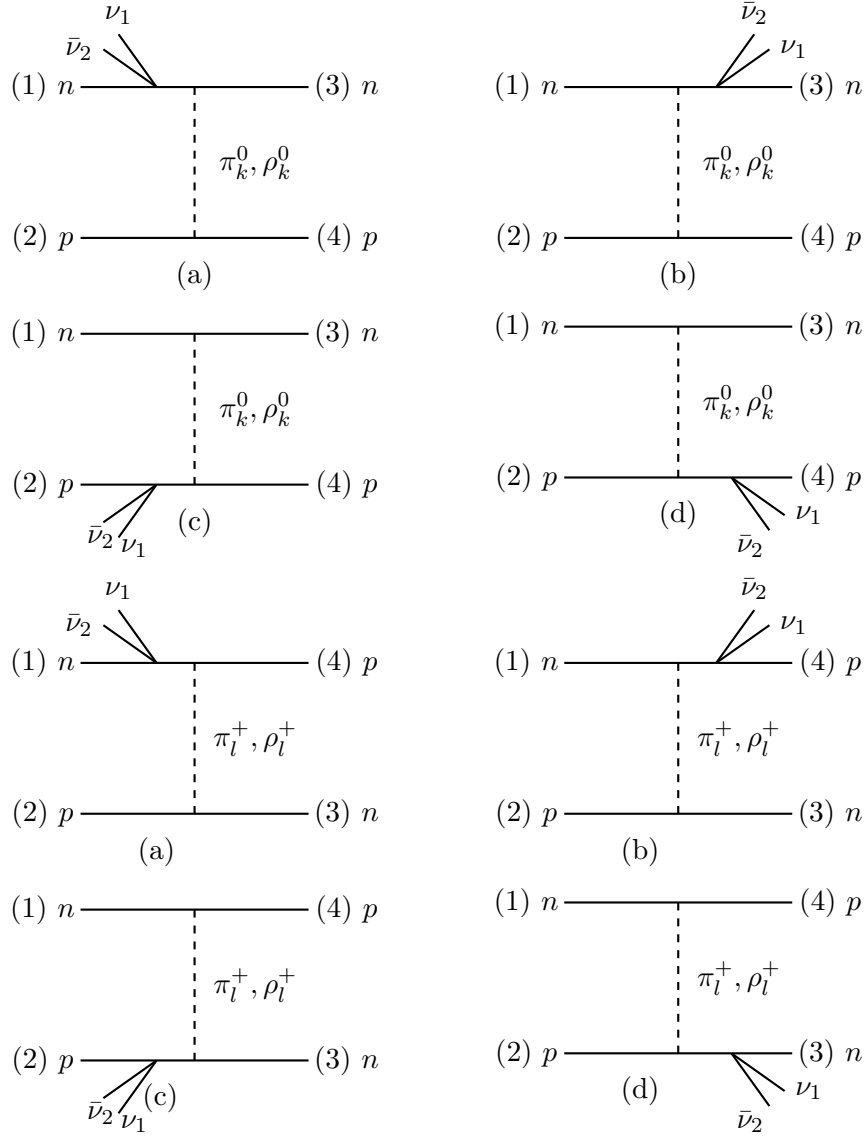

The structure of the calculation for neutron-proton bremsstrahlung is very similar to the previous one. After setting the symmetry factor $\mathcal{S}$ of Eq.~\ref{Eq:Rate} to one, instead of $1/4$, one needs to change the squared amplitude, which now reads
\begin{equation}
    \begin{split}
|\mathcal{M}|^2 =&~ 64\, G_{\rm F}^2\, g_A^2 \frac{\omega_1 \omega_2}{\omega^2} 
\Big[ h'^2_k +2h'^2_l+ 4 (h'_k-h'_l)(g'_k - g'_l+f'-g)  +6(g'_k - g'_l+f' - g)^2\\
& ~- 2 h'_l h'_k (1-(\hat{\bk}\cdot \hat{\bl})^2)\Big]
\\ \equiv&~ 64\, G_{\rm F}^2\, g_A^2 \frac{\omega_1\omega_2}{(\omega_1+\omega_2)^2}|m|^2,
    \end{split}
\end{equation}
where now we notice the presence of all Landau parameters (except for $f$, which does not contribute). The Feynman diagrams for this process are depicted in Fig.~\ref{DiagramNPBrem}; diagrams from $a)$ to $d)$ constitute the t-channel, where the exchange momentum is $\bk$, while diagrams from $e)$ to $h)$ are the u-channel, where the exchange momentum is $\bl$. We notice that our expression for the amplitude, even neglecting Landau terms, differs from that of \citetalias{Friman:1979ecl} for a factor of 2 missing in the third interference term.

Due to our assumed form for the interaction potential, the matrix element $|m|^2$ depends only on the modules of the momentum exchange $\bk$ in the $t$-channel and $\bl$ in the $u$-channel. Assuming $p_p\ll p_n$, the latter $\bl=\bp_4-\bp_1$ is dominated by the momentum of the neutron $-\bp_1$ on the Fermi surface, and $\bp_4$ is only a small correction. Since the matrix element itself $|m|^2$ is a slowly-varying function of $l=|\bl|$, we will expand $J$ to the first non-vanishing order in $\epsilon=p_p/p_n$ as

\begin{equation}\label{Eq:Jnp}
\begin{split}
    J^{\rm np}=&\frac{\pi}{2} \int_0^{2p_p} dk\left[|m|^2(k,p_n)+p_n\partial_l|m|^2(k,p_n)\frac{\epsilon^2}{4}   \left(1-\frac{3k^2}{4p_p^2}\right)+p_n^2\partial_l^2|m|^2(k,p_n)\frac{\epsilon^2}{4}\left(1-\frac{k^2}{4p_p^2}\right)\right]
    \end{split}
\end{equation}

\begin{figure}[htp!]
    \centering
    \includegraphics[scale = 0.6]{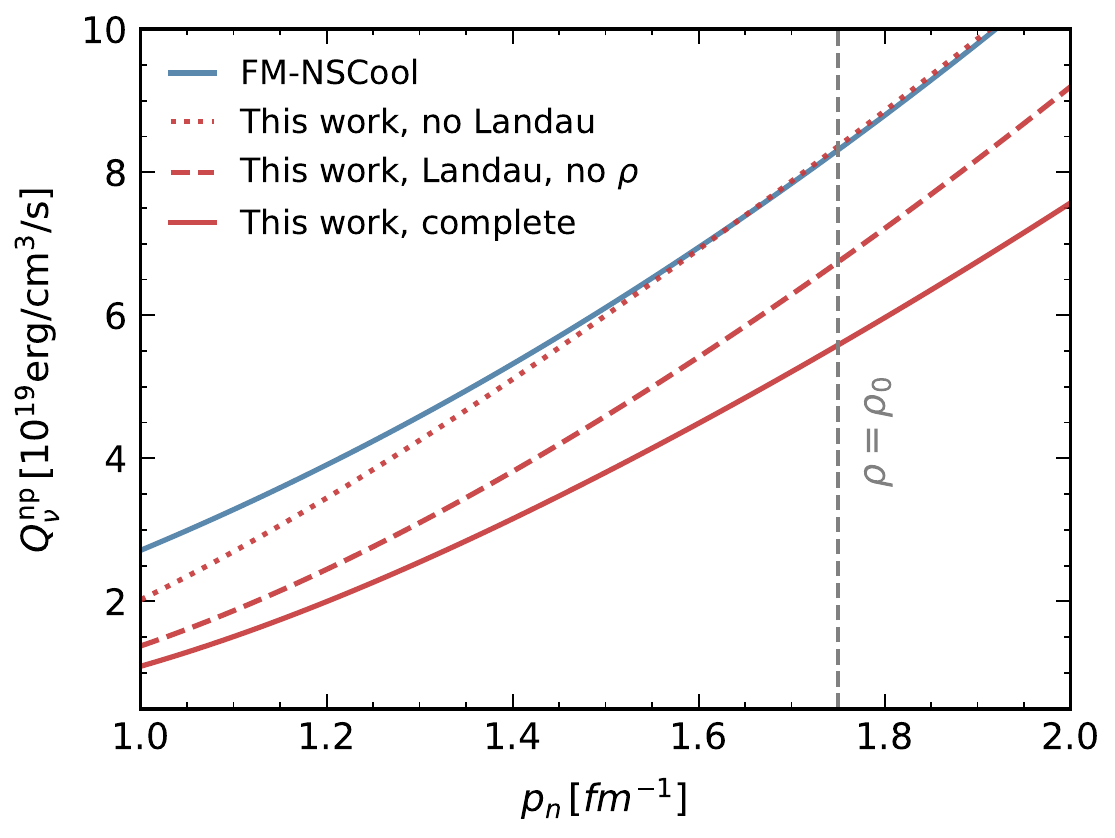}
    \caption{Comparison between the neutron-proton bremsstrahlung emissivity in this work (red curves) and in Eq.~(65b) of \citetalias{Friman:1979ecl} (blue curve) as a function of the neutron Fermi momentum. For all the curves we fixed $N_\nu =3$, $T = 10^9 \, \rm K$, $p_p = 85 \,  \Big(\frac{p_n}{340 \, \rm MeV}\Big)^2 \, \rm MeV$,we introduced the extra suppression factor ${\tt beta_{np} = 0.66}$ and we kept the nucleon masses fixed to their value in vacuum.}
    \label{fig:NP}
\end{figure}

We have checked that for typical values inside the core of the neutron star, keeping only the first term of the expansion leads to few \%. Putting all factors together we find
\begin{equation}\label{Eq:CompleteLnp}
Q_\nu^{\rm np}= \frac{41\, g_A^2G_{\rm F}^2m_n^2m_p^2 \,  T^8}{14175 \pi^{2}}J^{\rm np};
\end{equation}
keeping only the leading term in Eq.~\ref{Eq:Jnp} the final integration can be done analytically
\begin{equation}
\begin{split}
  J^{\rm np} =&\pi p_p\left[\frac{f_\pi^4}{m_\pi^4}\Big(2\eta^2(m_\pi)+4C_\rho^2\eta^2(m_\rho)+\phi(\alpha_p)+3C_\rho^2 \phi(\beta_p)+2C_\rho \Phi(\alpha_p,\beta_p)\right.\\
&-2\left(\eta(m_\pi)+C_\rho\eta(m_\rho)\right)\left(\psi(\alpha_p)+C_\rho \psi(\beta_p)\right)-4C_\rho^2\eta(m_\rho)\psi(\beta_p)\Big)+6(g-f')^2\\
  &\left.+4\frac{f_\pi^2}{m_\pi^2}(g-f')\left(\psi(\alpha_p)+2C_\rho\psi(\beta_p)-\eta(m_\pi)-2C_\rho\eta(m_\rho)\right)\right]\\
   &\equiv \pi p_p\mathcal{J}(\alpha_p, \beta_p),
\end{split}
\end{equation}

where $\alpha_p=m_\pi/(2p_p)$, $\beta_p=m_\rho/(2p_p)$, and $\eta(m)=p_n^2/(m^2+p_n^2)$. The functions $\phi(\alpha)$, $\Phi(\alpha,\beta)$ are defined in \eqref{eq:phi1} and \eqref{eq:phi2}, respectively, while $\psi(\alpha)$ is given by
\begin{align}
    \psi(\alpha)=1-\alpha\arctan\left(\frac{1}{\alpha}\right).
\end{align}
Again, this result must be multiplied by the number of neutrino flavors, $N_\nu = 3$, to obtain the \textit{total} luminosity. If we put $C_\rho$ and the Landau parameters to zero in our expression, and neglect the interference term $-2 \, \eta(m_\pi) \,\psi(\alpha_p)$, we recover Eq.~53a+53b of \citetalias{Friman:1979ecl} without Landau parameters. Our expression with the Landau parameters is different than that reported in \citetalias{Friman:1979ecl}, due to cancellations with the interference term, neglected in \citetalias{Friman:1979ecl}. In particular, we notice that the parameter $g'$ disappears completely. However, we have checked that this discrepancy is not the leading one in the final numerical difference between our results and the previous literature. 

In Fig.~\ref{fig:NP} we compare our complete emission rate (red curves) with the emission rates implemented in \href{http://www.astroscu.unam.mx/neutrones/NSCool/}{NSCool} (blue curve)
\begin{equation}\label{Eq:NSCoolNPBrem}
{\tt qbrem_{nn}=n_\nu \cdot 1.5d20 \cdot mstn(i)^2 mstp(i)^2 \cdot (kfp(i)/1.68d0) \cdot alpha_{np} \cdot beta_{np} \cdot (t/1.d9)^8},
\end{equation}
where ${\tt alpha_{np}=1.06, beta_{np} = 0.66}$ are kept constant in the code, ${\tt kfp(i)}$ is the proton Fermi momentum in units of $\rm fm^{-1}$,and ${\tt mstp(i)}$ represents the proton mass in units of the bare one. This expression coincides with Eq.~65b of \citetalias{Friman:1979ecl} and includes the contribution of the Landau parameters. The latter are seen to slightly decrease the emissivity, due to the negative contribution of the interference term. However, the values of these parameters, and therefore their impact on the emissivity, should not be taken as precise, given that they are only estimated for the case of nuclear-symmetric matter. Our analytical expressions allow to easily retrieve the emissivity for arbitrary values of the Landau parameters, directly assessing the impact of this uncertainty. In all curves we fixed also $N_\nu = 3$ for the number of neutrinos flavors, $T = 10^9 \, \rm K$, the nucleon masses to their in vacuum values, and the proton momentum to be $p_p = 85 \,  \Big(\frac{p_n}{340 \, \rm MeV}\Big)^2 \, \rm MeV$, following \citetalias{Friman:1979ecl}. 

The numerical discrepancy originates partially from the interference term $-2\eta(m_\pi)\psi(\alpha_p)$, not included in \citetalias{Friman:1979ecl} on the grounds that its enhancement to the emission would be approximately compensated by the $\rho$-meson exchange. However, we notice that the result in Eq.~70 of \citetalias{Friman:1979ecl} is already the \textit{full} result, while the authors seem to multiply it by an extra factor of 2 because of ``group II diagrams". It is unclear what these diagrams are, but it may be that this erroneous extra factor of 2 led to the claimed compensation with the $\rho$-meson effect. Consequently, for nuclear densities and above, the emission rate used in present numerical codes and most of the literature seems to be overestimated by a factor $\sim 2$. 

As a final remark, we note that the impact of the $\rho$ meson alone is less pronounced in this case compared to nucleon-nucleon bremsstrahlung. This is due to the presence of spin-spin interactions (Eq.~\ref{eq:gk}), which partially counteract the reduction in the tensor channel (Eq.~\ref{eq:hk}).

\subsection{MURCA}

We now pass to consider MURCA processes. Here we compute the rates of both the neutron branch
\begin{eqnarray}\label{Eq:MURCAProcess1}
n + n \rightarrow n + p + \ell + \bar{\nu}_{\ell}, \quad n + p + \ell \rightarrow n + n + \nu_{\ell}
\end{eqnarray}
and the proton branch
\begin{eqnarray}\label{Eq:MURCAProcess1p}
p + n \rightarrow p + p + \ell + \bar{\nu}_{\ell}, \quad p + p + \ell \rightarrow p + n + \nu_{\ell},
\end{eqnarray}
where $\ell=e,\mu$. The rate for direct and inverse process of neutrino emission are identical; this follows from the approximation that the neutrino energy is much smaller than the nucleon and electron energies, so that the hadronic matrix elements are independent of the neutrino energy, and from detailed balance. Thus, we introduce a factor of $2$ to account for both processes (we always consider the combined emission rate of neutrinos and antineutrinos; the two are of course identical by the same argument). We perform the computation for a generic lepton generation; the total emissivity in this case is then obtained by summing over the contributions of electrons and muons alone, since tau leptons are to0 heavy to be produced. This factor, however, is not a simple factor 2, since muons are not fully relativistic.

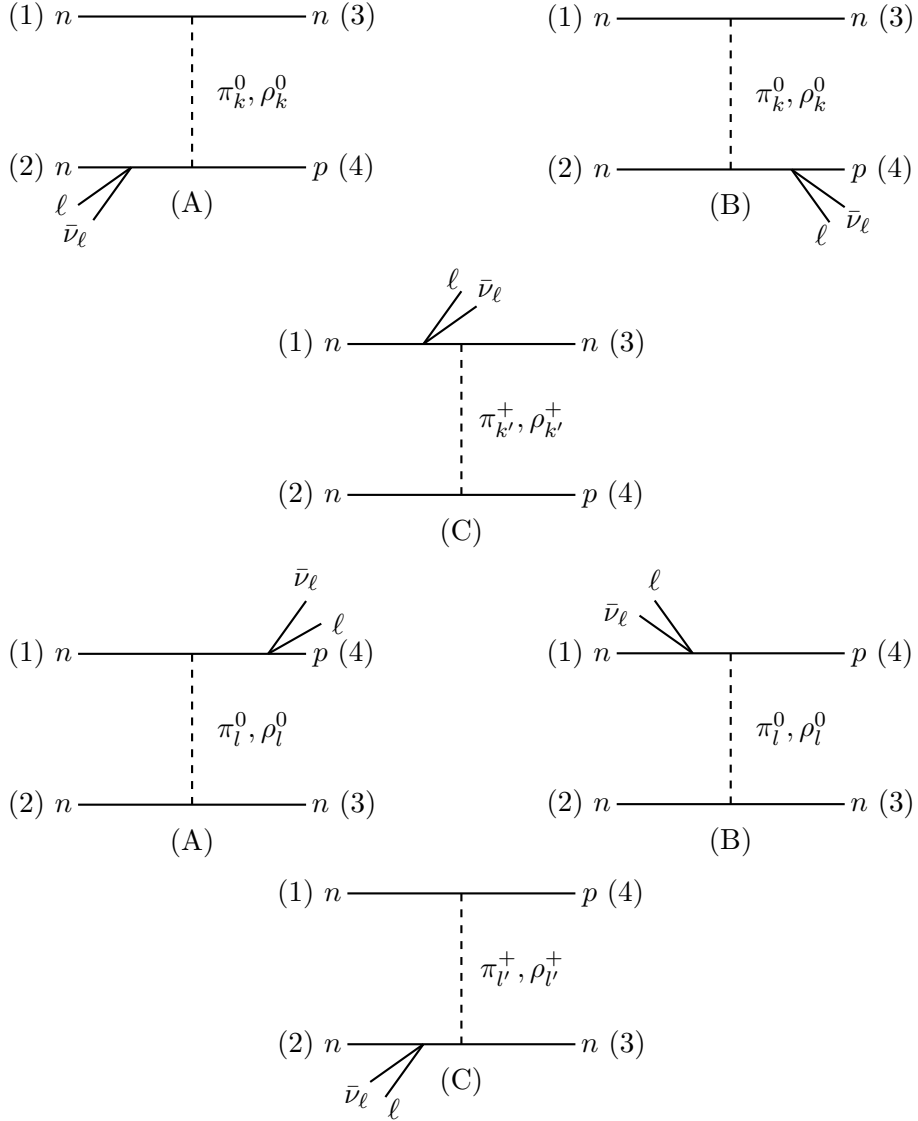
\begin{figure}[H]
    \centering
    \begin{minipage}{0.45\textwidth}
        \centering
        \begin{tikzpicture}[scale=1, transform shape]
            \node at (0,0) {(1) $n$};
            \node at (0,-2) {(2) $n$};
            \node at (4,0) {$n$ (3)};
            \node at (4,-2) {$p$ (4)};
            \draw[thick] (0.5,0) -- (3.5,0);
            \draw[thick] (0.5,-2) -- (3.5,-2);
            \draw[thick, dashed] (2,0) -- (2,-2);
            \node at (2.8,-1) {$\pi^0_k, \rho^0_k$};
            \draw[thick] (1.2,-2) -- ++(-0.7,-0.5) node [left] {$\ell$};
            \draw[thick] (1.2,-2) -- ++(-0.5,-0.7) node [below left=-2pt] {$\bar{\nu}_\ell$};
            \node at (2,-2.5) {(A)};
        \end{tikzpicture}
    \end{minipage}
    \begin{minipage}{0.45\textwidth}
        \centering
        \begin{tikzpicture}[scale=1, transform shape]
            \node at (0,0) {(1) $n$};
            \node at (0,-2) {(2) $n$};
            \node at (4,0) {$n$ (3)};
            \node at (4,-2) {$p$ (4)};
            \draw[thick] (0.5,0) -- (3.5,0);
            \draw[thick] (0.5,-2) -- (3.5,-2);
            \draw[thick, dashed] (2,0) -- (2,-2);
            \node at (2.8,-1) {$\pi^0_k, \rho^0_k$};
            \draw[thick] (2.8,-2) -- ++(0.7,-0.5) node [below left=2.5pt] {$\ell$};
            \draw[thick] (2.8,-2) -- ++(0.5,-0.7) node [right=2pt] {$\bar{\nu}_\ell$};
            \node at (2,-2.5) {(B)};
        \end{tikzpicture}
    \end{minipage}
    \begin{minipage}{0.9\textwidth}
        \centering
        \begin{tikzpicture}[scale=1, transform shape]
            \node at (0,0) {(1) $n$};
            \node at (0,-2) {(2) $n$};
            \node at (4,0) {$n$ (3)};
            \node at (4,-2) {$p$ (4)};
            \draw[thick] (0.5,0) -- (3.5,0);
            \draw[thick] (0.5,-2) -- (3.5,-2);
            \draw[thick, dashed] (2,0) -- (2,-2);
            \node at (2.8,-1) {$\pi^+_{k'}, \rho^+_{k'}$};
            \draw[thick] (1.5,0) -- ++(0.7,0.5) node [above left=2.5pt] {$\ell$};
            \draw[thick] (1.5,0) -- ++(0.5,0.7) node [right=2pt] {$\bar{\nu}_\ell$};
            \node at (2,-2.5) {(C)};
        \end{tikzpicture}
    \end{minipage}

    \begin{minipage}{0.45\textwidth}
        \centering
        \begin{tikzpicture}[scale=1, transform shape]
            \node at (0,0) {(1) $n$};
            \node at (0,-2) {(2) $n$};
            \node at (4,0) {$p$ (4)};
            \node at (4,-2) {$n$ (3)};
            \draw[thick] (0.5,0) -- (3.5,0);
            \draw[thick] (0.5,-2) -- (3.5,-2);
            \draw[thick, dashed] (2,0) -- (2,-2);
            \node at (2.8,-1) {$\pi^0_l, \rho^0_l$};
            \draw[thick] (3,0) -- ++(0.7,0.4) node [right] {$\ell$};
            \draw[thick] (3,0) -- ++(0.5,0.7) node [above ] {$\bar{\nu}_\ell$};
            \node at (2,-2.5) {(A)};
        \end{tikzpicture}
    \end{minipage}
    \begin{minipage}{0.45\textwidth}
        \centering
        \begin{tikzpicture}[scale=1, transform shape]
            \node at (0,0) {(1) $n$};
            \node at (0,-2) {(2) $n$};
            \node at (4,0) {$p$ (4)};
            \node at (4,-2) {$n$ (3)};
            \draw[thick] (0.5,0) -- (3.5,0);
            \draw[thick] (0.5,-2) -- (3.5,-2);
            \draw[thick, dashed] (2,0) -- (2,-2);
            \node at (2.8,-1) {$\pi^0_l, \rho^0_l$};
            \draw[thick] (1.5,0) -- ++(-0.7,0.5) node [left] {$\bar{\nu}_\ell$};
            \draw[thick] (1.5,0) -- ++(-0.5,0.7) node [above] {$\ell$};
            \node at (2,-2.5) {(B)};
        \end{tikzpicture}
    \end{minipage}
    \begin{minipage}{0.9\textwidth}
        \centering
        \begin{tikzpicture}[scale=1, transform shape]
            \node at (0,0) {(1) $n$};
            \node at (0,-2) {(2) $n$};
            \node at (4,0) {$p$ (4)};
            \node at (4,-2) {$n$ (3)};
            \draw[thick] (0.5,0) -- (3.5,0);
            \draw[thick] (0.5,-2) -- (3.5,-2);
            \draw[thick, dashed] (2,0) -- (2,-2);
            \node at (2.8,-1) {$\pi^+_{l'}, \rho^+_{l'}$};
            \draw[thick] (1.5,-2) -- ++(-0.7,-0.5) node [below right=2.5pt] {$\ell$};
            \draw[thick] (1.5,-2) -- ++(-0.5,-0.7) node [left=2pt] {$\bar{\nu}_\ell$};
            \node at (2,-2.5) {(C)};
        \end{tikzpicture}
    \end{minipage}

    \caption{Feynman diagram for the MURCA processes in the neutron branch. Diagrams a-c constitute the t-channel, while d-f represent the u-channel.}
    \label{DiagramMURCA}
\end{figure}

In this case we also find it useful to first compute the neutrino \textit{emission rate} and then the final emissivity. The two are easily related as
\begin{equation}
Q_\nu^{\rm MURCA} = 2\int \frac{d^3 p_\nu}{(2\pi)^3} \omega_\nu \Gamma_\nu^{\rm MURCA};
\end{equation}
where we remind the factor 2 for the direct and inverse processes. This is particularly convenient because with the neutrino emission rate at hand, one can also easily obtain the \textit{neutrino absorption rate}, another important quantity in the physics of neutron stars.

\subsubsection{Neutron branch}

We start studying the neutron branch, which is supposed to be the most relevant one for typical equations of state~\cite{Yakovlev:2000jp}. To make contact with basically all previous literature, we first perform the computation in the triangular approximation, i.e neglecting protons and leptons momenta $p_n \gg p_p ,p_\ell$ in the Dirac-$\delta$ for the conservation of momentum. We then proceed to perform the computation dropping this approximation, and quantifying the discrepancy.  

\paragraph{Triangular approximation}

The squared amplitude for the neutron branch MURCA process $n(\bp_1)+n(\bp_2)\rightarrow n(\bp_3)+p(\bp_4)+\ell(\bp_\ell)+\bar{\nu}(\bp_\nu)$, in the triangular approximation is
\begin{equation}
\begin{split}
|\mathcal{M}|^2&=64G^2g_A^2 \frac{\omega_\nu\omega_\ell}{(\omega_\nu+\omega_\ell)^2}\left[12(f'-g)^2+\frac{21}{4}\left(\frac{f_\pi}{m_\pi}\right)^4\frac{p_n^4}{(p_n^2+m_\pi^2)^2}\left(1-C_\rho\frac{p_n^2+m_\pi^2}{p_n^2+m_\rho^2}\right)^2\right]\\
&\equiv 64\, G^2g_A^2\frac{\omega_\nu}{\mu_\ell}|m|^2\ .
\end{split}
\end{equation}
where $\mu_\ell$ is the Fermi energy and where we have assumed that $p_n\gg p_p,p_e$, so that $\bk\approx-\bp_2$, $\bl\approx -\bp_1$, and $\bk\cdot\bl=- p_n^2/2$. Notice that even for mildly relativistic or non-relativistic muons, the leptonic trace still leads to the characteristic $\omega_\nu \omega_\ell$ product in the numerator, since any term proportional to the momentum of the particle averages to zero due to isotropy of the emitted neutrino. In the last step we also used the fact that electrons and muons are degenerate in the NS core, with $\omega_\ell \sim \mu_\ell \gg \omega_\nu\approx T$. We highlight that when computing the complete amplitude squared, all the Landau parameters but $f'$ and $g$ drop out in the triangular approximation. This was not appreciated in \citetalias{Friman:1979ecl}, where in the computation of the ``exchange terms" Landau parameters were not included. 

We now compute the emission rate
\begin{equation}
\begin{split}
\Gamma^{\rm M, T}_{\omega_\nu}=&\frac{1}{2}\frac{1}{2\omega_\nu}\int\frac{d^3 p_1}{(2\pi)^3}\int\frac{d^3 p_2}{(2\pi)^3}\int\frac{d^3 p_3}{(2\pi)^3}\int\frac{d^3 p_4}{(2\pi)^3}\int\frac{d^3 p_\ell}{(2\pi)^3(2 \, \omega_\ell)}64G^2g_A^2\frac{\omega_\nu}{\mu_\ell}|m|^2\times \\
&\times (2\pi)^4\delta(E_1+E_2-E_3-E_4-\omega_\ell-\omega_\nu)\delta^{(3)}(\bp_1+\bp_2-\bp_3-\bp_4-\bp_\ell-\bp_\nu)\times\\
&\times f_1f_2(1-f_3)(1-f_4)(1-f_\ell),
\end{split}
 \end{equation}
where the superscript ``M,T" stands for ``MURCA, Triangular",
and the initial factor $1/2$ is the symmetry factor for identical particles in the initial state. Using the fact that $|m|^2$ is constant and the degeneracy of the nucleons, then
\begin{equation}
\begin{split}
\Gamma^{\rm M, T}_{\omega_\nu}=&G^2g_A^2\frac{8m_n^3m_pp_n^3p_p|m|^2}{ (2\pi)^{11}}\frac{p_\ell}{\mu_\ell}\int\mathrm{d}\Omega_l\prod_{i=1}^4\mathrm{d}\Omega_i\delta^{(3)}(\bp_1+\bp_2-\bp_3-\bp_4-\bp_\ell-\bp_\nu)\times\\
    &\times \int\mathrm{d}\omega_e\prod_{i=1}^4\mathrm{d}E_i\delta(E_1+E_2-E_3-E_4-\omega_\ell-\omega_\nu)f_1f_2(1-f_3)(1-f_4)(1-f_\ell)\\
    \equiv &G^2g_A^2\frac{8m_n^3m_pp_n^3p_p|m|^2}{ (2\pi)^{11}} \frac{p_\ell}{\mu_\ell} \,\mathcal{A}\mathcal{E},
\end{split}
\end{equation}
where $p_\ell$ if the Fermi momentum of the emitted lepton (either an electron or a muon). The integral over the energies can be easily performed in the complex plane and gives
\begin{equation}
\mathcal{E}=\frac{1}{24}\frac{(\omega_\nu^2+9\pi^2T^2)(\omega_\nu^2+\pi^2T^2)}{e^{\omega_\nu/T}+1},
\end{equation}
while the angular integral (under the triangular approximation) reads
\begin{equation}
    \mathcal{A}=\frac{128\pi^4}{p_n^3}.
\end{equation}
Putting everything together we get
\begin{equation}
\begin{split}
\Gamma^{\rm M, T}_{\omega_\nu}&=\frac{8G^2g_A^2p_pm_n^3m_p|m|^2}{3(2\pi)^7}\frac{(\omega_\nu^2+9\pi^2T^2)(\omega_\nu^2+\pi^2T^2)}{e^{\omega_\nu/T}+1}\\&=\frac{8G^2g_A^2m_n^3m_pp_p}{3(2\pi)^7}\frac{(\omega_\nu^2+9\pi^2T^2)(\omega_\nu^2+\pi^2T^2)}{e^{\omega_\nu/T}+1}\frac{p_\ell}{\mu_\ell}\\
&\times\left[12(f'-g)^2+\frac{21}{4}\left(\frac{f_\pi}{m_\pi}\right)^4\frac{p_n^4}{(p_n^2+m_\pi^2)^2}\left(1-C_\rho\frac{p_n^2+m_\pi^2}{p_n^2+m_\rho^2}\right)^2\right]
\end{split}
\end{equation}

Finally, the total energy loss rate is 
\begin{equation}\label{Eq:FinalMURCAtriangular}
\begin{split}
\mathcal{Q}^{\rm M, T}_\nu =&2 \left(1+\frac{p_{F\mu}}{\mu_\mu}\right)\times
\frac{11513}{120960\pi}\frac{G^2g_A^2f_\pi^4}{m_\pi^4}m_n^3m_pp_pT^8\times \\
&\times \frac{1}{2}\times\frac{21}{16}\times\left[\frac{32}{7}\left(\frac{m_\pi}{f_\pi}\right)^4(f'-g)^2+2\left(\eta(m_\pi)-C_\rho\eta(m_\rho)\right)^2\right] \\ \equiv & 2 \left(1+\frac{p_{F\mu}}{\mu_\mu}\right)\times
\frac{11513}{120960\pi}\frac{G^2g_A^2f_\pi^4}{m_\pi^4}m_n^3m_pp_pT^8\alpha_{\rm MURCA},
\end{split}
\end{equation}
where the factor of 2 in front of everything takes into account the inverse reaction, $n + p + l \rightarrow n + n + \nu_{\ell}$, where we assumed the electrons to be fully relativistic (and therefore $p_e \simeq \mu_e$), and where
\begin{equation}\label{eq:alphaMurca}
   \alpha_{\rm MURCA}= \frac{1}{2}\times\frac{21}{16}\times\left[\frac{32}{7}\left(\frac{m_\pi}{f_\pi}\right)^4(f'-g)^2+2\left(\eta(m_\pi)-C_\rho\eta(m_\rho)\right)^2\right]\ .
\end{equation}

\begin{figure}[H]
    \centering
    \includegraphics[scale = 0.55]{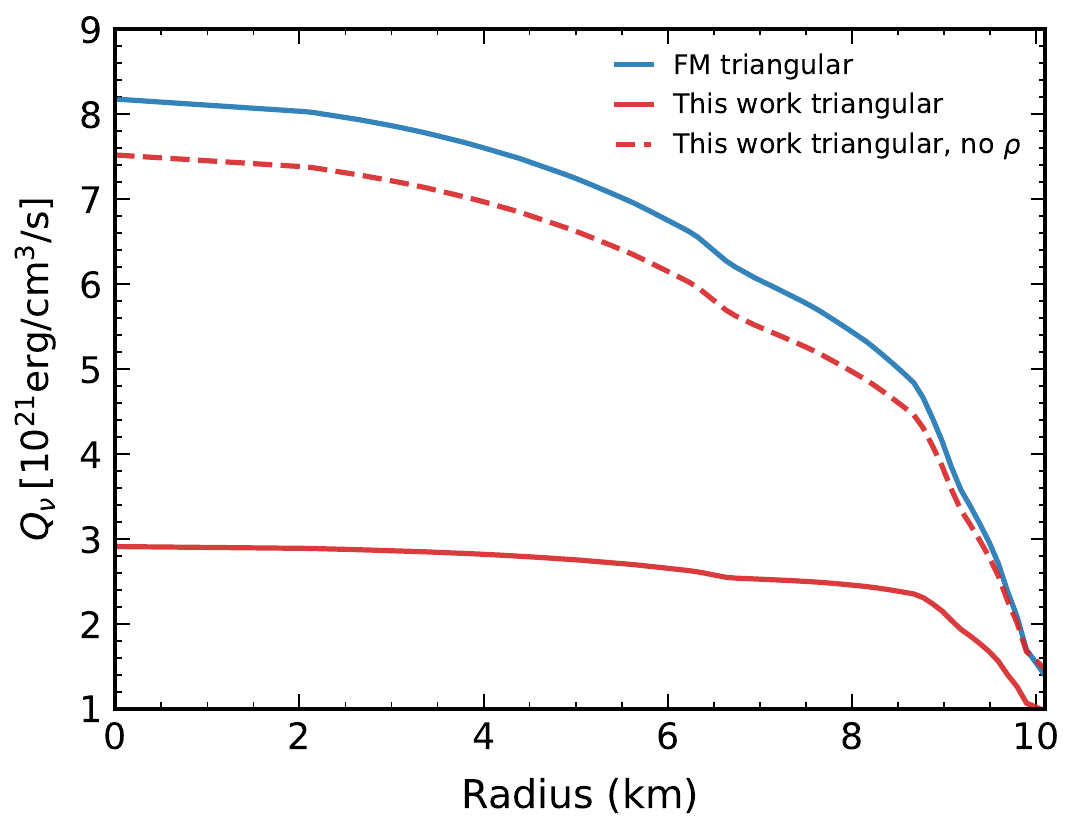}
    \caption{MURCA emissivity in a realistic NS profile with APR EOS using our Eq.~\ref{Eq:FinalMURCAtriangular} (red curves) with (solid) and without (dashed) $\rho$-meson exchange, and the expression employed in \href{http://www.astroscu.unam.mx/neutrones/NSCool/}{NSCool} (blue curve). In both cases the temperature was fixed to $T = 10^9 \, \rm K$, the nucleon masses to their bare values, the Landau parameters to the values quoted in \citetalias{Friman:1979ecl} and we fixed the short-range physics suppression factor to ${\tt beta_n  =0.68d0}$. }
    \label{Fig:MURCA_triangular}
\end{figure}

Compared to Eq. 56 of \citetalias{Friman:1979ecl}, in addition to the factor 21/16 coming from the addition of the u-channel and interference terms, we get a different structure of the Landau parameters and a factor of 1/2 in the OPE plus $\rho$-exchange terms. We have been able to reproduce the amplitude squared in Eq.~39 of \citetalias{Friman:1979ecl}, including Landau parameters, which is the sum of t-channel and u-channel amplitudes squared (the two are the same). However, we have not been able to trace back the factor of 2 discrepancy in their final emission rate, nor the meaning of the so called ``exchange diagrams" in this case, which would correspond to our u-channel diagrams. The inconsistency extends to their Eq. 75, where the ratio between emissivity with and without "exchange terms" should be $\sim 0.65$, not $\sim 1.3$, based on their own rates. Nevertheless, it seems that all the subsequent literature~\cite{Yakovlev:2000jp, Yakovlev1995A&A, Kopp:2023sev} has adopted this erroneous factor of two in the emission rate. Then, the inclusion of the $\rho$-meson further suppresses the emissivity, which has been therefore overestimated by more than a factor of 2 at nuclear densities.

The MURCA process is typically the most important for NS cooling. Given its particular relevance, in this case to make the comparison between our findings and previous results, we use a realistic NS profile provided in \href{http://www.astroscu.unam.mx/neutrones/NSCool/}{NSCool} for a NS of $1~ M_\odot$ with the Akmal-Pandharipande-Ravenhall (APR) equation of state (EOS)~\cite{Akmal:1998cf} . In Fig.~\ref{Fig:MURCA_triangular} we show the comparison for this NS radial profile between our result for the triangular approximation (red curve), Eq.~\ref{Eq:FinalMURCAtriangular}, and the one reported in \href{http://www.astroscu.unam.mx/neutrones/NSCool/}{NSCool} (blue curve) again in units of $\rm erg/cm^3/s$
\begin{equation}
{\tt Murca_n = 8.55d21 \cdot mstn(i)^3 \cdot mstp(i) \cdot (kfe(i) + kfm(i))/1.68d0 \cdot alpha_n \cdot beta_n \cdot (t/1.d9)^8}\nonumber, 
\end{equation}
with ${\tt alpha_n =1.76d0-0.63d0 \cdot (1.68d0/kfn(i))^2}$ and ${\tt beta_n  =0.68d0}$. This equation coincides with Eq.~65b of \citetalias{Friman:1979ecl}, with the addition of the muon contribution. Apart from the differences mentioned above, we notice two further issues with this expression: the presence of the Fermi momentum of the electron, $p_e$, rather than the one of the proton; the muon contribution is overestimated because it doesn't take into account that, contrary to electrons, muons are not relativistic or mildly relativistic. These two problems were already noticed recently in Ref.~\cite{Kopp:2023sev}. However, we stress that these corrections alone tend to make the output of \href{http://www.astroscu.unam.mx/neutrones/NSCool/}{NSCool} deviate further from our exact computation.

Fig.~\ref{Fig:MURCA_triangular} makes manifest that the results presented in the literature overestimate the MURCA emission rate by more than a factor of 2. The biggest impact comes from the $\rho$-meson exchange, whose contribution is not compensated by other factors when everything is computed self-consistently. We verified that instead the impact of the Landau parameters is modest.

\paragraph{General expression}

We now want to check the impact of the triangular approximation; as a simple reason for doubting its full applicability, we note that if the proton and electron momenta are aligned their sum would be essentially identical to the neutron Fermi momentum. 

Two things need to be changed: the squared amplitude and then the corresponding \textit{angular} integration, while the rest stays the same. The squared amplitude $|\mathcal{M}|^2$ can be written as usual in terms of a reduced one  $|\mathcal{M}|^2 = 64G^2\frac{\omega_\nu}{\omega_l} |m|^2$. In turn, the reduced amplitude squared can be written as $|m|^2 = |m_u|^2+ |m_t|^2 + 2  \Re(m_u^* m_t)$, where the first two terms are the u and t channels amplitude squared, while the third one is their interference. With the usual momentum assignments $n(\bp_1)+n(\bp_2)\rightarrow n(\bp_3)+p(\bp_4)+\ell(\bp_\ell)+\bar{\nu}(\bp_\nu)$, the t-channel amplitude squared is written as follows
\begin{equation}
\begin{split}  
|m_t|^2 =& g_A^2\left[6(f'^2+g^2)-2\left(3f'g'_k+3(f'+2g)g'_{\tilde{k}}+f'h'_k+(f'+2g)h'_{\tilde{k}}\right)+\right.\\
&+\left.(3g'^2_k+9g'^2_{\tilde{k}}+2g'_kh'_k+h'^2_k+6g'_{\tilde{k}}h'_{\tilde{k}}+3h'^2_{\tilde{k}})\right]+3(g'_k-g'_{\tilde{k}})^2+2(g'_k-g'_{\tilde{k}})(h'_k-h'_{\tilde{k}})\\
&+(h'_k-h'_{\tilde{k}})^2+2h'_kh'_{\tilde{k}}\left(1-(\hat{\bk}\cdot\hat{\tilde{\bk}})^2\right)
\end{split}    
\end{equation}
where $\bk = \bp_1 - \bp_3$ and $\tilde{\bk} = \bp_2 - \bp_4$. The u-channel piece, $|m_u|^2$, is the same but with $\bk \rightarrow \bl = \bp_2 -  \bp_3$ and $\tilde{\bk} \rightarrow \tilde{\bl} = \bp_1 -  \bp_4$. The interference term reads instead
\begin{equation}
    \begin{split}
       2  \Re &(m_u^* m_t) = g_A^2\left[-24 f' g + 
 2 \left(2 f' \left(3( g'_{\tilde{k}} + g'_{\tilde{l}} )+ h'_{\tilde{k}} + h'_{\tilde{l}}\right) + 
    3 g(g'_{k} +  g'_{\tilde{k}} + g'_{l} + g'_{\tilde{l}})\right.\right.\\
    &+\left. g(h'_{k} + h'_{\tilde{k}} + 
       h'_{l} + h'_{\tilde{l}})\right)-\left(g'_{l} (h'_{k} + h'_{\tilde{k}}) +
    g'_{\tilde{l}} (h'_{k} + 5 h'_{\tilde{k}}) +h'_{k} h'_{l} - h'_{\tilde{k}} h'_{l} -
    h'_{k} h'_{\tilde{l}} + 3 h'_{\tilde{k}} h'_{\tilde{l}} +\right.\\
    &+\left.
    g'_{k} (3 g'_{l} + 3 g'_{\tilde{l}} + h'_{l} + h'_{\tilde{l}}) +
    g'_{\tilde{k}} (3 g'_{l} + 15 g'_{\tilde{l}} + h'_{l} + 5 h'_{\tilde{l}})\right)+\\
    &+\left. h'_{k}  h'_{l} \left(1 - (\hat{\bk}\cdot\hat{\bl})^2\right) - 
 2 h'_{\tilde{k}}  h'_{l} \left(1 - (\hat{\tilde{\bk}}\cdot\hat{\bl})^2\right) - 
 2 h'_{k}  h'_{\tilde{l}} \left(1 - (\hat{\bk}\cdot\hat{\tilde{\bl}})^2\right) + 
 2 h'_{\tilde{k}}  h'_{\tilde{l}} \left(1 - (\hat{\tilde{\bk}}\cdot\hat{\tilde{\bl}})\right)^2\right]+\\
 &+\left[(g'_l-g'_{\tilde{l}})(h'_k-h'_{\tilde{k}})+(g'_k-g'_{\tilde{k}})(h'_l-h'_{\tilde{l}})-(h'_k-h'_{\tilde{k}})(h'_l-h'_{\tilde{l}})+3(g'_l-g'_{\tilde{l}})(g'_k-g'_{\tilde{k}})+\right.\\
 &+\left.2 h'_{k}  h'_{l} \left(1 - (\hat{\bk}\cdot\hat{\bl})^2\right) - 
 2 h'_{\tilde{k}}  h'_{l} \left(1 - (\hat{\tilde{\bk}}\cdot\hat{\bl})^2\right) - 
 2 h'_{k}  h'_{\tilde{l}} \left(1 - (\hat{\bk}\cdot\hat{\tilde{\bl}})^2\right) + 
 2 h'_{\tilde{k}}  h'_{\tilde{l}} \left(1 - (\hat{\tilde{\bk}}\cdot\hat{\tilde{\bl}})^2\right)\right]
    \end{split}
\end{equation}
where $x_{ij}$, with $i, j = k, \tilde{k}, l, \tilde{l}$ is the cosine of the angle between the corresponding vectors. The emission rate is then
\begin{equation}
\begin{split}
&\Gamma^{\rm MURCA,n}_{\omega_\nu}=\frac{8G_{\rm F}^2p_n^3p_pp_lm_n^3m_p}{(2\pi)^{11}\mu_\ell} \mathcal{E}\int\mathrm{d}\Omega_\ell\prod_{i=1}^4\mathrm{d}\Omega_i\delta^{(3)}(\bp_1+\bp_2-\bp_3-\bp_4-\bp_\ell-\bp_\nu)|m|^2 \nonumber \\
&= \frac{256G_{\rm F}^2p_n^3p_pp_l m_n^3m_p\, \mathcal{E}}{(2\pi)^{11}\mu_\ell} \frac{1}{p_n^3 p_p p_\ell} \int d^3p_\ell\int \prod_{i=1}^4 d^3 p_i \delta(p_i^2 - p_{F,i}^2) \delta^3(\bp_1 + \bp_2 - \bp_3 - \bp_4 - \textbf{p}_\ell) |m|^2 \nonumber \\ 
&= \frac{256G_{\rm F}^2m_n^3m_p\mathcal{E}}{(2\pi)^{11} \mu_\ell}\int \prod_{i=1}^4 d^3 p_i \delta(p_i^2 - p_{F,i}^2) |m|^2,
\end{split}
\end{equation}

where in the last step we used the spatial $\delta$ of Dirac to integrate the electron momentum. 

At this point, in order to perform the angular integrations more easily, we find it useful to perform the following change of basis (with unit determinant) $\bp_1, \bp_2, \bp_3, \bp_4 \rightarrow \bp_2, \bl, \btl, \bp_4$. With this transformation, the rate reads
\begin{equation}\label{Eq:MURCA_rate_general_1}
\begin{split}
\Gamma^{\rm MURCA,n}_{\omega_\nu}&=\frac{256 \, G^2m_n^3m_p\mathcal{E}}{ (2\pi)^{11} \mu_\ell}\underbrace{\int d^3 p_2 \delta(p_2^2 - p_n^2)}_{\frac{p_n}{2}\int d\Omega_2} \underbrace{\int d^3 p_4 \delta(p_4^2 - p_p^2)}_{\frac{p_p}{2}\int d\Omega_4} \int d^3 l d^3 \tilde{l} \delta(|\bl-\bp_2|^2-p_n^2)\times \nonumber \\ 
&\times \delta(|\btl+\bp_4|^2-p_n^2)\delta(p_\ell^2-|\bl+\btl|^2)|m|^2 =  \\
&=\frac{16 \, G^2m_n^3m_p\mathcal{E}}{(2\pi)^{9}\mu_\ell} \int_{\Sigma} dld\tilde{l}\int_0^{2\pi} d\phi \int_0^{2\pi} d\beta |m|^2,
\end{split}
\end{equation}
where the region $\Sigma$ is defined by
\begin{equation}
\Sigma:~|\tilde{l}-p_n|<p_p,\quad 0<l<2p_n,\quad l+\tilde{l}>p_\ell,\quad |l-\tilde{l}|<p_\ell
\end{equation}

where it is understood that all the scalar products fixed by the delta functions must be consistently replaced in the reduced squared amplitude. The angles $\phi$ and $\beta$ are the azimuthal angles compared to the plane containing $\bl$ and $\tilde{\bl}$ of $\bp_2$ and $\bp_4$, respectively. They are defined through the relations  
\begin{equation}
        \sin\phi=\frac{\bp_2\cdot(\bl\times\btl)}{p_nl\tilde{l}\sqrt{1-x_{2l}^2}\sqrt{1-x_{l\tilde{l}}^2}}, \quad \sin\beta=\frac{\bp_4\cdot(\bl\times\btl)}{p_pl\tilde{l}\sqrt{1-x_{4\tilde{l}}^2}\sqrt{1-x_{l\tilde{l}}^2}},
\end{equation}
where $x_{2l}=\bp_2\cdot\bl/(p_n l)$ and $x_{4\tilde{l}}=\bp_4\cdot \btl/(p_p \tilde{l})$ are fixed by the kinematic constraints in \eqref{Eq:MURCA_rate_general_1}.


\begin{figure*}[ht]
\vskip-5pt
\includegraphics[width=0.5\columnwidth]{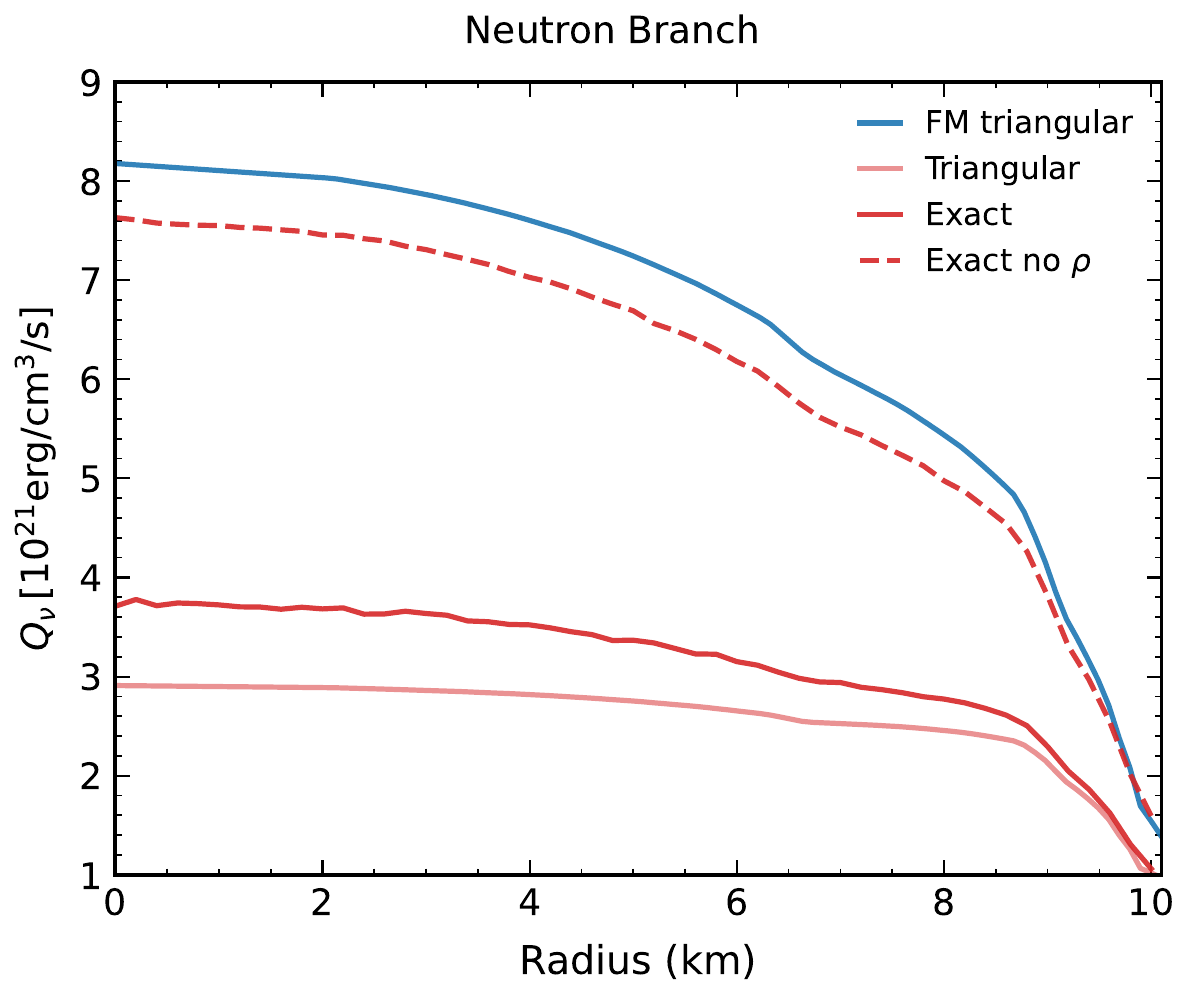}
\includegraphics[width=0.49\columnwidth]{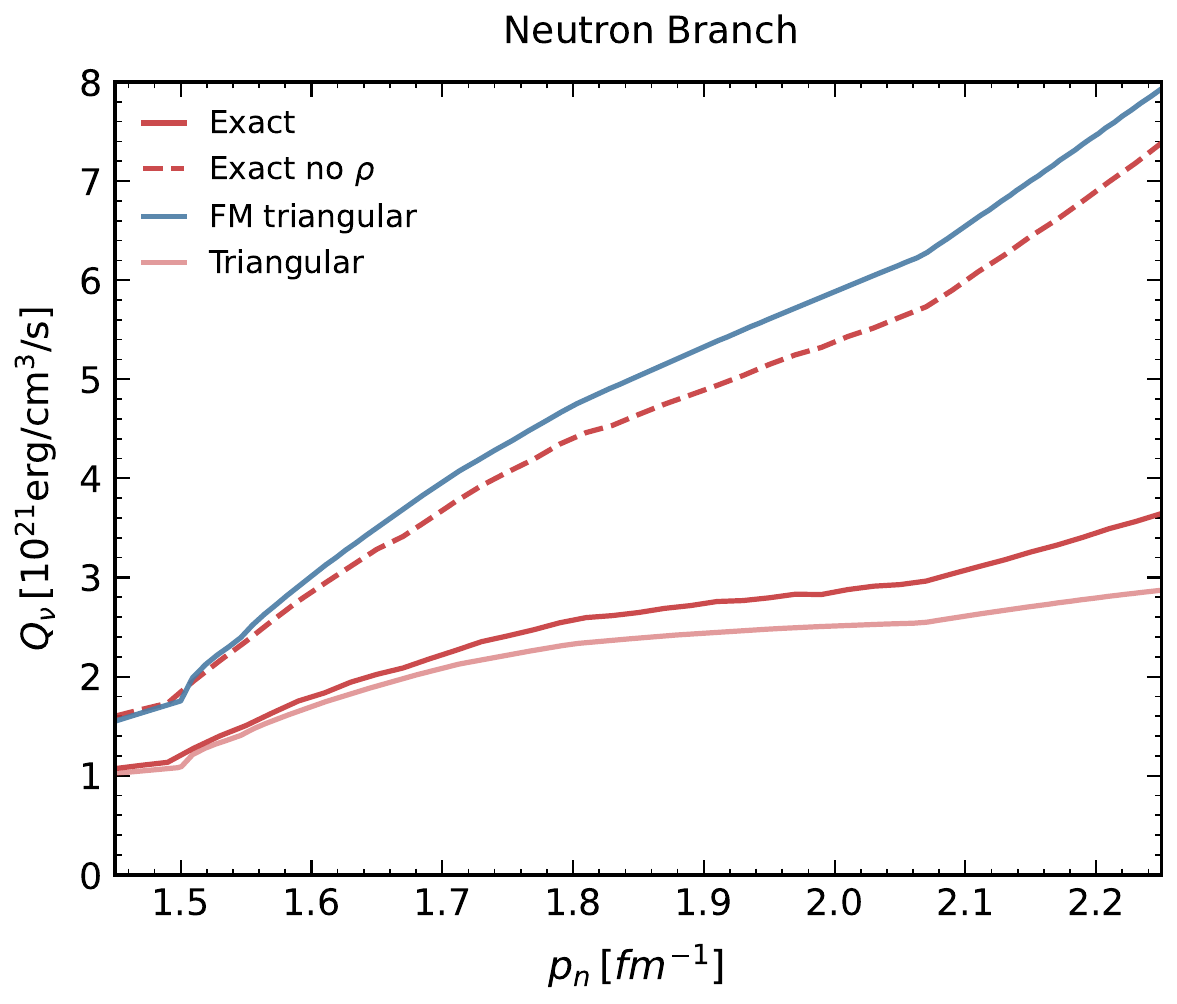}
\vskip-5pt
\caption{Exact MURCA emissivity for our Eq.~\ref{eq:MURCAfull} with (solid red curves) and without (dashed) $\rho$-meson exchange, compared again with the expression employed in \href{http://www.astroscu.unam.mx/neutrones/NSCool/}{NSCool} (blue curve) and our triangular approximation (shaded red curve). We see that the impact of the triangular approximation is around $\sim 20-30 \%$. In all cases the temperature was fixed to $T = 10^9 \, \rm K$, the nucleon masses to their bare values, the Landau parameters to the values quoted in \citetalias{Friman:1979ecl} and we fixed the short-range physics suppression factor to ${\tt beta_n  =0.68d0}$. The left panel shows the emissivity as a function of the radius of one solar mass NS with APR EOS. In the right hand panel we show the same quantities but as a function of the neutron momentum in the NS determined from the same profile used in the left panel.}
\label{Fig:MURCA_fullN}
\vskip-10pt
\end{figure*}

The remaining four integrals must be done numerically after having written the reduced amplitude squared in this new convenient basis. Finally, the emissivity is given by

\begin{equation}\label{eq:MURCAfull}
    Q_\nu^{\rm MURCA, n}=\frac{11513}{483840}\frac{G^2m_n^3m_p}{(2\pi)^{3}\mu_\ell} \int_{\Sigma} dld\tilde{l}\int_0^{2\pi} d\phi \int_0^{2\pi} d\beta |m|^2.
\end{equation}

We show our final results for Eq.~\ref{eq:MURCAfull} in Fig.~\ref{Fig:MURCA_fullN} with (solid red) and without (dashed) $\rho$-meson exchange. Compared to our triangular approximation (shaded red curve), the full numerical results differ by a factor $\sim 20-30 \%$ at most. Somewhat surprisingly -- given that in the considered NS model $p_p + p_e \sim p_n$ -- the triangular approximation provides quite precise results. In all cases the temperature was fixed to $T = 10^9 \, \rm K$, the nucleon masses to their bare values, the Landau parameters to the values quoted in \citetalias{Friman:1979ecl} and we fixed the short-range physics suppression factor to ${\tt beta_n  =0.68d0}$.

\begin{figure*}[ht]
\vskip 5pt
\includegraphics[width=0.5\columnwidth]{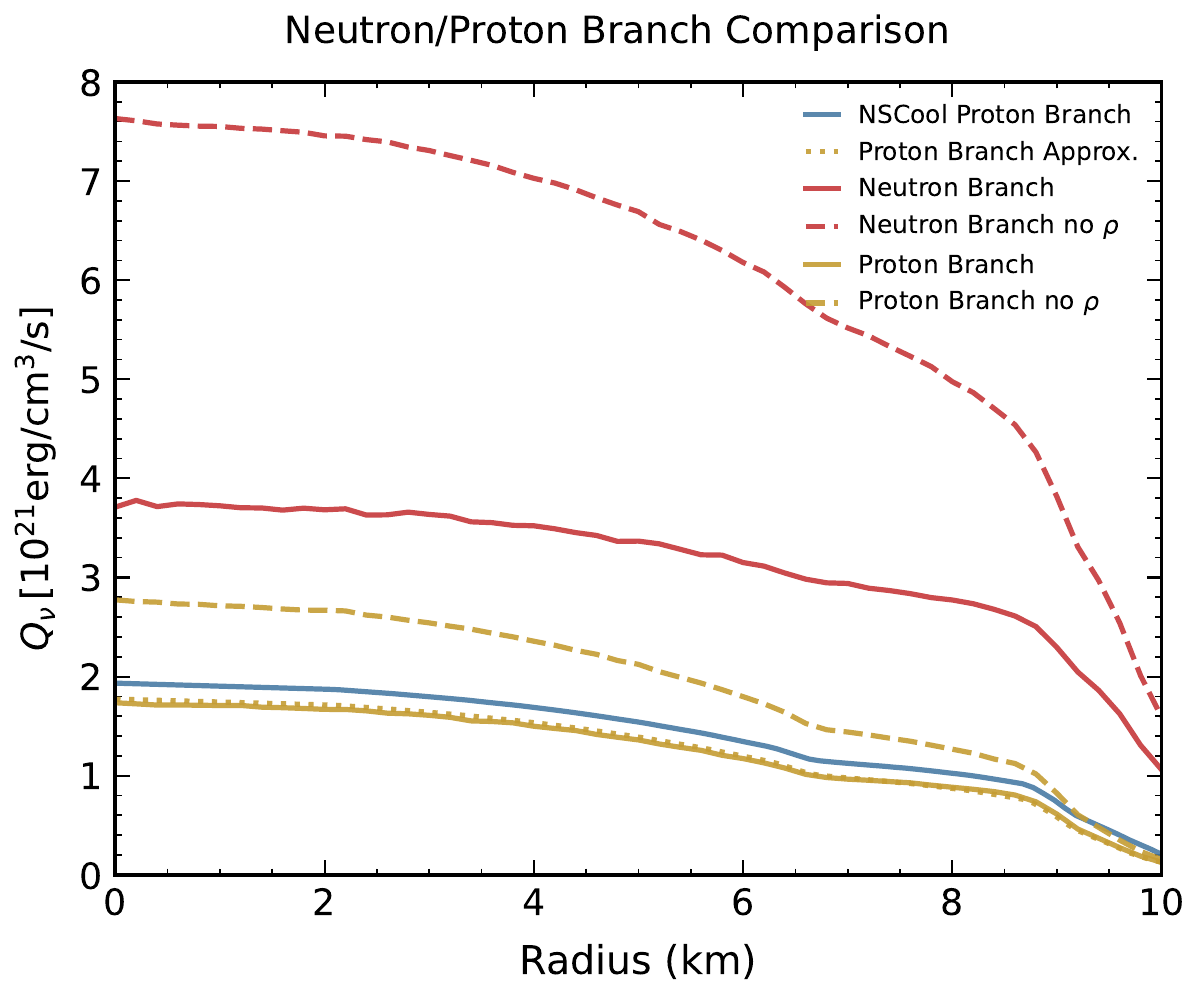}
\includegraphics[width=0.49\columnwidth]{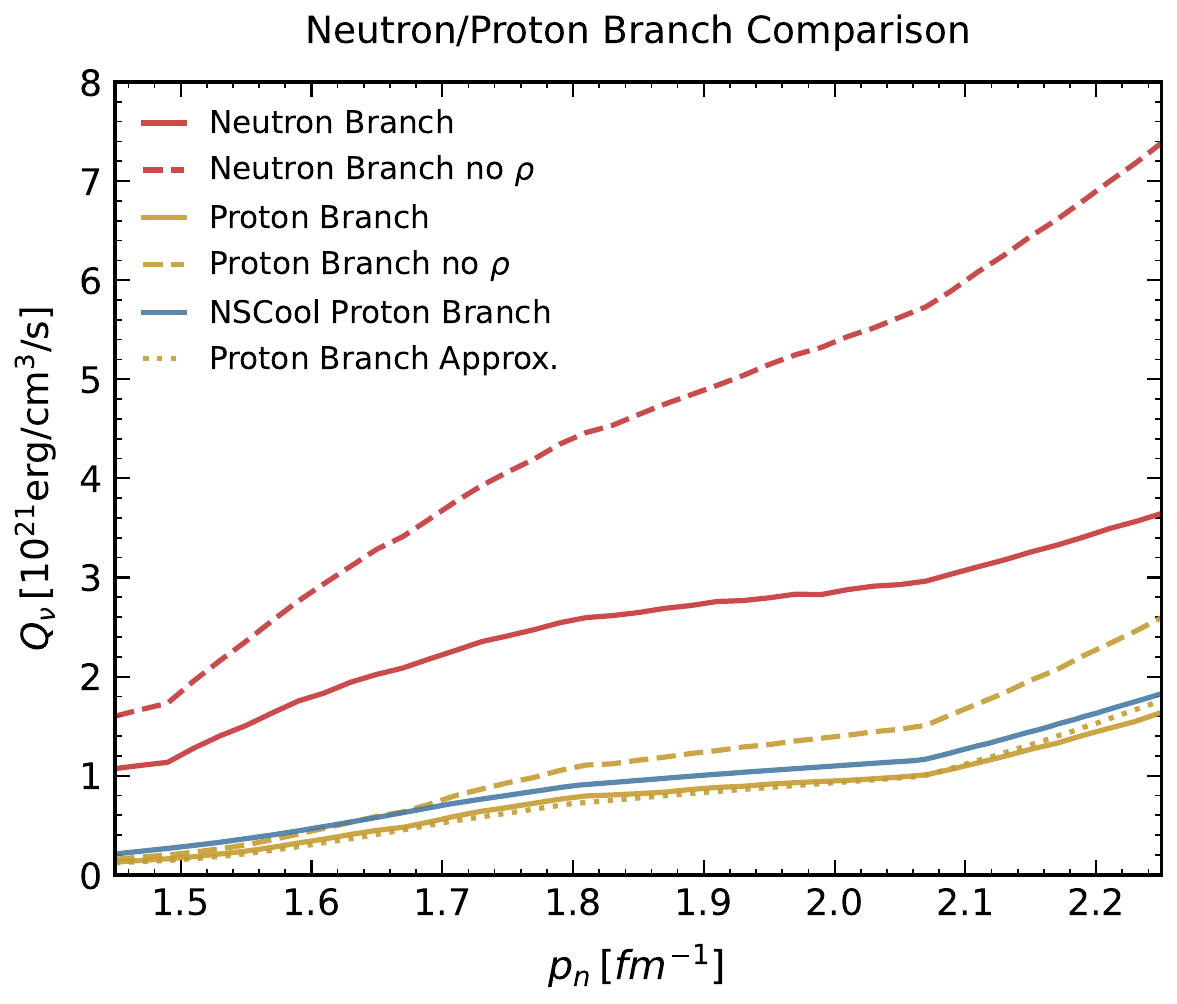}
\vskip-5pt
\caption{Exact MURCA emissivity for the proton branch in Eq.~\ref{eq:MURCAfullP} (yellow curves), compared with the neutron branch result in Eq.~\ref{eq:MURCAfull} (red curve) and with the proton branch expression employed in \href{http://www.astroscu.unam.mx/neutrones/NSCool/}{NSCool} (blue curve). We see that the proton branch is always subdominant but not by a large amount, although it will generally be more suppressed once effective masses are properly taken into account (because the effective proton mass in NS is usually smaller than the effective neutron mass). Again in all cases the temperature was fixed to $T = 10^9 \, \rm K$, the nucleon masses to their bare values, the Landau parameters to the values quoted in \citetalias{Friman:1979ecl} and we fixed the short-range physics suppression factor to ${\tt beta_n  =0.68d0}$. As in Fig.~\ref{Fig:MURCA_fullN} the left panel shows the emissivity as a function of the radius of one solar mass NS with APR EOS, while the right hand panel shows the same quantities but as a function of the neutron momentum in the NS determined from the same profile used in the left panel.}
\label{Fig:MURCA_fullNPcomparison}
\vskip-10pt
\end{figure*}

\subsubsection{Proton branch}

The results described in the previous sections are for the MURCA neutron branch. Now we provide the results also for the proton branch and check its relevance. Assuming the following momentum assignments $p(\bp_1)+n(\bp_2)\rightarrow p(\bp_3)+p(\bp_4)+\ell(\bp_\ell)+\bar{\nu}(\bp_\nu)$ and based on crossing-symmetry arguments, it is easy to realize that the squared amplitude for the proton branch can be obtained by replacing $\bk\rightarrow -\bk$, $\btk\rightarrow -\btk$, and $\bl\leftrightarrow \btl$. Thus we can immediately write the neutrino emission rate as
\begin{equation}
\Gamma^{\rm MURCA,p}_{\omega_\nu}=\frac{16 G^2m_nm_p^3\mathcal{E}}{(2\pi)^{9}\mu_\ell} \theta(p_l+3p_p-p_n)\int_{\Sigma} dld\tilde{l}\int_0^{2\pi} d\phi \int_0^{2\pi} d\beta |m|^2,
\end{equation}
where the angles $\phi$ and $\beta$ are defined as before while the region $\Sigma$ now is given by
\begin{equation}
\Sigma:~|l-p_n|<p_p,\quad 0<\tilde{l}<2p_p,\quad l+\tilde{l}>p_\ell,\quad |l-\tilde{l}|<p_\ell,
\end{equation}
and finally
\begin{equation}\label{eq:MURCAfullP}
    Q_\nu^{\rm MURCA, p}=\frac{11513}{483840}\frac{G^2m_p^3m_n}{(2\pi)^{3}\mu_\ell} \theta(p_l+3p_p-p_n)\int_{\Sigma} dld\tilde{l}\int_0^{2\pi} d\phi \int_0^{2\pi} d\beta |m|^2.
\end{equation}

We show the result for the proton branch (yellow curves), compared with the neutron branches (red curves) in Fig.~\ref{Fig:MURCA_fullNPcomparison}, where we also show the proton branch implemented in \href{http://www.astroscu.unam.mx/neutrones/NSCool/}{NSCool} (blue curve)
\begin{eqnarray}
{\tt qmurca_p=8.55d21 \cdot mstn(i) \cdot mstp(i)^3 \cdot 
(kfe(i)/1.68d0)\cdot (kfe(i)+} \nonumber\\ + {\tt 3.d0*kfp(i)-kfn(i))^2/(8.d0 \cdot kfe(i) \cdot kfp(i)) \cdot alpha_p \cdot beta_p \cdot (t/1.d9)^8},\nonumber \\ 
\, \, \,
\end{eqnarray}
where ${\tt alpha_p, beta_p}$ are the same as for the neutron branch. The proton branch is always subdominant but not by a large amount, although this may change when effective masses are properly taken into account, because the effective proton mass in NS is usually smaller than the effective neutron mass. 


For completeness, in Fig.~\ref{Fig:MURCA_fullNPcomparison} we also show the results in an approximation analogous (dotted red) to the triangular one for the neutron branch. In the proton branch case this approximation amounts to take $k$ and $l$ to be the maximum exchanged momenta~\cite{Yakovlev:2000jp}
\begin{equation}
\begin{split}
\mathcal{Q}^{\rm MT, p}_\nu =&2 \times \left(\sum_{\ell=e,\mu}\frac{(p_\ell+3p_p-p_n)^2}{\mu_\ell}\right)\times
\frac{11513}{645120\pi}G^2g_A^2m_n^3m_pT^8\times \\
&\times \left[2(f'-g)^2+\left(\frac{f_\pi}{m_\pi}\right)^4\frac{(p_n-p_p)^4}{((p_n-p_p)^2+m_\pi^2)^2}\left(1-C_\rho\frac{(p_n-p_p)^2+m_\pi^2}{(p_n-p_p)^2+m_\rho^2}\right)^2\right].
\end{split}
\end{equation}
This result -- which we label ``Proton Branch Approx." in our figure -- agrees very precisely with the output of \href{http://www.astroscu.unam.mx/neutrones/NSCool/}{NSCool}, and it is also very similar to the exact computation.

\section{Neutrino absorption rate}

As MURCA and its inverse is the dominant process of neutrino emission, their time-reversed versions
\begin{equation}
    n + n + \nu_{\ell} \rightarrow n + p + \ell,\quad n+p+\ell+\overline{\nu}_\ell\to n+n
\end{equation}
are among the primary absorption mechanisms in neutron star matter (the primary one for old neutron stars). The absorption rates can be directly related to the emission ones from the principle of detailed balance; once the emission rate $\Gamma^{\rm em}_\nu(\omega_\nu)$ for a neutrino with energy $\omega_\nu$ is known, the absorption rate $\Gamma^{\rm abs}_\nu(\omega_\nu)$ can be found as
\begin{equation}
    \Gamma^{\rm abs}_\nu(\omega_\nu)=\Gamma^{\rm em}_\nu(\omega_\nu) e^{\omega_\nu/T}.
\end{equation}
Notice that there is a somewhat conventional choice in the definition of absorption rate; in a given environment, the evolution of the neutrino distribution is driven by the full collisional term
\begin{equation}
    \left(\frac{\partial f_\nu}{\partial t}\right)_{\rm coll}=\Gamma^{\rm em}_\nu(\omega_\nu)(1-f_\nu)-\Gamma^{\rm abs}_\nu(\omega_\nu) f_\nu,
\end{equation}
which can be rewritten as
\begin{equation}
    \left(\frac{\partial f_\nu}{\partial t}\right)_{\rm coll}=\Gamma^{\rm em}_\nu(\omega_\nu)-\tilde{\Gamma}^{\rm abs}_\nu(\omega_\nu) f_\nu,
\end{equation}
where we defined an enhanced absorption rate
\begin{equation}
    \tilde{\Gamma}^{\rm abs}_\nu(\omega_\nu)=\Gamma^{\rm em}_\nu(\omega_\nu)+\Gamma^{\rm abs}_\nu(\omega_\nu)=\Gamma^{\rm em}_\nu(\omega_\nu) (1+e^{\omega_\nu/T})
\end{equation}
which also accounts for the suppression in the inverse emission process caused by the Pauli exclusion principle. This is simply called the absorption rate in \citetalias{Friman:1979ecl}; here for clarity we will adopt the more conventional definition $\Gamma^{\rm abs}_\nu(\omega_\nu)$.

\begin{figure}[H]
    \centering
    \includegraphics[scale = 0.5]{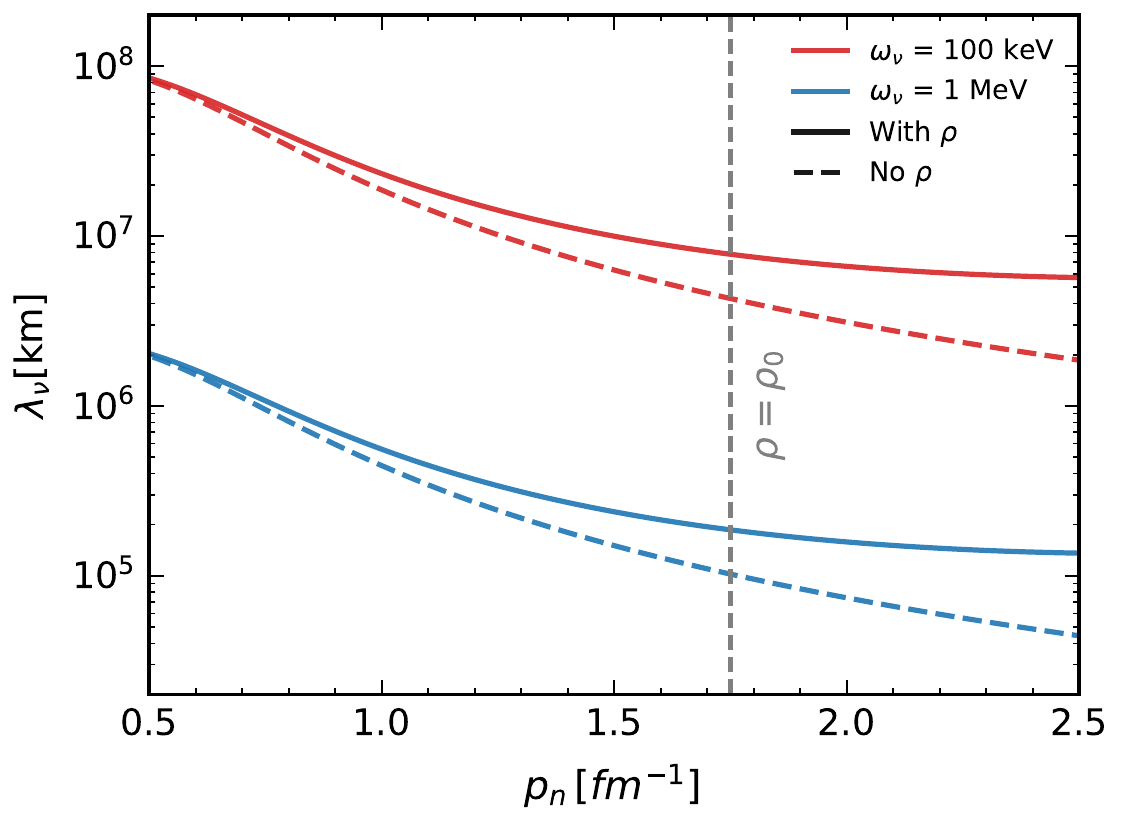}
    \caption{Electron neutrino MFP, $\lambda_\nu = 1/\Gamma^{\rm abs}_\nu$, as a function of the neutron Fermi momentum for $\omega = T = 100 \, \rm keV$ (red curves) and $\omega = T = 1 \, \rm MeV$ (blue curves)  with (solid) or without (dashed) the contribution from $\rho$ exchange. For this plot we also fixed $p_p = 85 \,  \Big(\frac{p_n}{340 \, \rm MeV}\Big)^2 \, \rm MeV$, $T = 10^9 \, \rm K$, bare nucleon masses and standard values for the Landau parameters. }
    \label{Fig:MFP_MURCA}
\end{figure}

Thus, in the triangular approximation the absorption rate (for the neutron branch, which is the dominant one) is easily found to be
\begin{equation}\label{Eq:GammaAbs}
\begin{split}
\Gamma^{\rm abs}_{\omega_\nu}&=\frac{2G^2g_A^2m_n^3m_pp_p}{(2\pi)^7}\frac{(\omega_\nu^2+9\pi^2T^2)(\omega_\nu^2+\pi^2T^2)}{e^{-\omega_\nu/T}+1}\frac{p_\ell}{\mu_\ell}\\
&\times\left[16(f'-g)^2+\left(\frac{f_\pi}{m_\pi}\right)^4\frac{7p_n^4}{(p_n^2+m_\pi^2)^2}\left(1-C_\rho\frac{p_n^2+m_\pi^2}{p_n^2+m_\rho^2}\right)^2\right],
\end{split}
\end{equation}
which also corresponds to the emission rate with the substitution $\omega_\nu \rightarrow - \omega_\nu$. In Fig.~\ref{Fig:MFP_MURCA} we show the neutrino mean free path (MFP) for $\omega = T = 100 \, \rm keV$ (red curves) and $\omega = T = 1 \, \rm MeV$ (blue curves)  with (solid) or without (dashed) the contribution from $\rho$ exchange. We also fixed $p_p = 85 \,  \Big(\frac{p_n}{340 \, \rm MeV}\Big)^2 \, \rm MeV$, $T = 10^9 \, \rm K$, bare nucleon masses and standard values for the Landau parameters. The MFP exceeds by several orders of magnitude the typical radius of a NS for both cases, with or without the inclusion of a $\rho$ meson. We also checked that the impact of Landau parameters in these conditions is very modest.

\section{Conclusions}

The cooling of NSs via neutrino emission plays a fundamental role in their evolution, especially in its early stage. A clear understanding of its quantitative impact is therefore necessary to compare with the evolution of surface temperature and luminosity of isolated NSs. Obviously this process is directly affected by multiple uncertainties from nuclear physics, especially in regards to how the nucleon-nucleon interaction is modeled, both in vacuum and in the dense nuclear medium. However, the large nuclear physics uncertainties should not obscure the importance of the particle physics framework and approximations used to obtain the cooling rates. If anything, the existence of already large uncertainties in the nuclear physics sector should push us further into clarifying completely the particle physics aspects -- i.e., phase-space and matrix element evaluation -- of this process.

In this work, we have moved from this push into a complete reevaluation of the cooling rates from $nn$ bremsstrahlung, $np$ bremsstrahlung, and MURCA processes. For all three processes, we have found a wide range of differences compared to the seminal treatment in~\citetalias{Friman:1979ecl}. Generally, these differences seem to come mainly from the counting of different groups of diagrams, and the neglect of interference diagrams for certain processes. We have also folded in the suppression of the nucleon interaction potential at large momentum exchange, modeled as a rho-meson exchange, which was in principle present in~\citetalias{Friman:1979ecl} but argued to cancel with the interference diagrams. We do not find evidence for this cancellation, especially in the cooling rate as a function of density; rather, neglecting the rho-meson exchange provides an additional cause for overestimation of the cooling rates. Further, we have gone beyond the conventional triangular approximation, in which the Fermi momenta of protons and electrons are neglected compared to the ones of neutrons; while by itself this seems to be a 20-30\% effect, piled up with the remaining differences it conspires to create significant discrepancies with the results from previous literature. We refer to our main text for a detailed discussion of each of the differences in treatment, and their impact, for each of the processes. Here we rather focus on the implications for the phenomenological studies of neutron star cooling.

\begin{figure}[H]
    \centering
    \includegraphics[scale = 0.55]{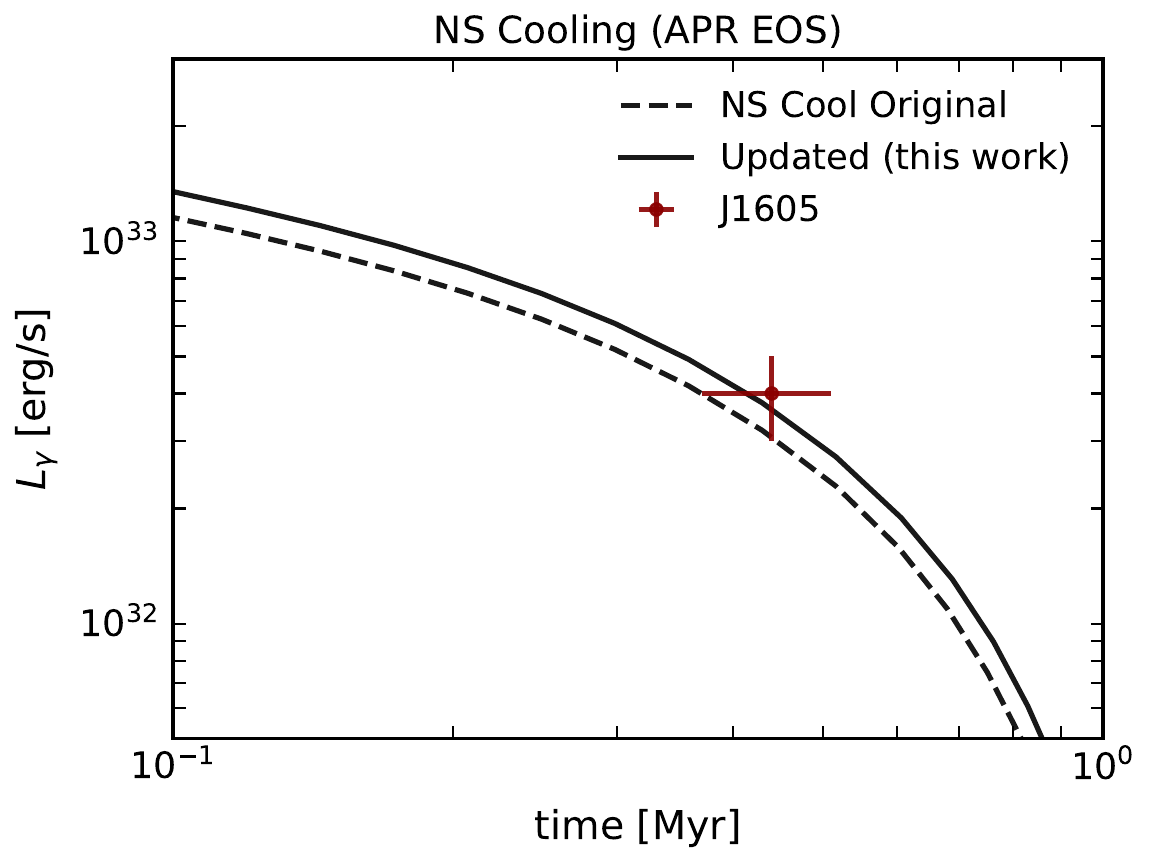}
    \caption{Cooling curve for a NS of mass $M_{\rm NS} = M_\odot$ with APR EOS with (solid) and without (dashed) taking into account the corrections derived in this work.}
    \label{Fig:NSCooling}
\end{figure}

We find that the cooling rates for MURCA, the dominant process, can be even a factor $2$ lower when integrated over the entire star, and even more for the space-dependent emission rate per unit volume. For the bremsstrahlung processes, which are subdominant, similar discrepancies are found. These differences are evaluated using the same short-range interactions as~\citetalias{Friman:1979ecl}, modeled with Landau parameters which are however highly uncertain. Our analytical treatment is particularly suitable to evaluate the impact of these uncertainties, and flexible enough to be adapted to specific choices of the Landau parameters. These discrepancies come entirely from particle physics aspects; to maintain a consistent and fair comparison, we have stuck to the same framework as~\citetalias{Friman:1979ecl} for the modeling of the nucleon-nucleon interaction potential. A more refined treatment of this point would certainly be of interest, but we notice that NS cooling is anyway at present generally described following~\citetalias{Friman:1979ecl}.

The impact of these corrections on the full evolution of NSs due to their cooling needs to be assessed. To illustrate the significance of these order-one factors, we present in Fig.~\ref{Fig:NSCooling} the time evolution of the luminosity for a NS of one solar mass, with APR EOS and without superfluidity, computed using NSCool. The dashed curve represents the result using the incorrect MURCA rate, while the solid curve shows the result with our corrected rate. We compare these theoretical curves with the properties of the isolated neutron star RX J1605.3~\cite{Tetzlaff_2012, Pires:2019qsk}. Notably, with the old rates, the APR EOS almost fails to match the observations within their error bars. However, using the correct MURCA emissivity leads to an excellent fit. We stress again that we did not attempt to tackle in-medium corrections, which may well be even more important than the various discrepancies we outlined in this work. Nevertheless, some of the matter effects such as the effective nucleon masses or the quenching of the axial couplings, are typically implemented as an overall rescaling of the ~\citetalias{Friman:1979ecl} framework we adopted.

Intriguingly, the numerical corrections we have derived can have an important impact beyond purely astrophysical questions. A precision treatment of NS cooling, compared with the time evolution of isolated NSs, is also the basis of powerful constraints on novel particle emission, and especially for the QCD axion~\cite{Caputo:2024oqc, Buschmann:2021juv, Sedrakian:2015krq, Hamaguchi:2018oqw, IwamotoPhysRevD.64.043002, Leinson:2021ety, Leinson:2019cqv}. In addition, the axion emission is evaluated in the same framework as the neutrino one, and therefore must also be reassessed. Thus, axion constraints would be directly affected by the results of this work. We will proceed with a reassessment of the bounds on the QCD axion in a forthcoming work~\cite{AxionFuture}. 

\section*{Acknowledgments}
We wish to thank Georg Raffelt and Toby Opferkuch for detailed comments on the manuscript, Bengt L. Friman for instructive email exchanges about nucleon effective masses, and Joachim Kopp and Edoardo Vitagliano for helpful discussions. AC also thanks Stefan Stelzl for enlightening discussions on related topics. SB is supported by the Israel Academy of Sciences and Humanities \& Council for Higher Education Excellence Fellowship Program for International Postdoctoral
Researcher. DFGF is supported by the Villum Fonden under Project No. 29388 and the European Union’s Horizon 2020 Research and Innovation Program under the Marie Sklodowska-Curie Grant Agreement No. 847523 “INTERACTIONS.”

\appendix

\section{Explicit MURCA computation}

Given the confusion in the literature and the various order one factors we corrected, we find it useful to report here an explicit computation. In particular, we proceed in computing everything in a non-relativistic framework; however, for the pion and $\rho$-meson exchange we also checked our results using a relativistic framework, computing the amplitude squared with {\tt FeynCalc~9.3.1} \cite{Mertig:1990an, Shtabovenko:2016sxi, Shtabovenko:2020gxv} and then performing a non-relativistic expansion of the final results. In this appendix we limit ourselves to MURCA, which is in any case the most relevant process for NS cooling. 

We start by considering a single Landau parameter as a simple jumping off point. Therefore, we take the following effective non-relativistic potential for nucleon-nucleon interaction
\begin{equation*}\label{Eq:PotentialG}
V_g(\mathbf{k}) = g \, \boldsymbol{\sigma}_1 \cdot \boldsymbol{\sigma}_2,
\end{equation*}
and we start computing the amplitude for the t-channel from the diagrams $a)$ to $c)$ in Fig.~\ref{DiagramMURCA}. For the diagram $a)$ the amplitude reads
\begin{equation}
\mathcal{M}_a = \frac{iGg}{\sqrt{2}} \frac{\ell^\mu}{\omega} \chi_3^\dagger \sigma^j \chi_1 \chi_4^\dagger \tau^- (\delta_{\mu 0} - g_A \delta_{\mu i} \sigma_i) \sigma^j \chi_2,
\end{equation}
where in isospin space we have $\chi_4^\dagger = \begin{pmatrix}
0 & 1
\end{pmatrix}$, $\chi_2 = \begin{pmatrix}
1 \\ 
0
\end{pmatrix}$ and
$\tau^- = \begin{pmatrix}
0 & 0 \\
1 & 0
\end{pmatrix}$. So in this case the isospin structure gives just a trivial factor $\chi_4^\dagger \tau^- \chi_2 = 1$, so that the amplitude is simply
\begin{equation}
\mathcal{M}_a = \frac{iGg}{\sqrt{2}} \frac{\ell^\mu}{\omega} \chi_3^\dagger \sigma^j \chi_1 \chi_4^\dagger (\delta_{\mu 0} - g_A \delta_{\mu i} \sigma_i) \sigma^j \chi_2.
\end{equation}

Analogously for the diagram $b)$ we have 
\begin{equation}
\mathcal{M}_b = \frac{iGg}{\sqrt{2}} \Big(-\frac{\ell^\mu}{\omega}\Big) \chi_3^\dagger \sigma^j \chi_1 \chi_4^\dagger \sigma^j(\delta_{\mu 0} - g_A \delta_{\mu i} \sigma_i) \chi_2,
\end{equation}
where the minus sign of difference comes from the nucleon propagator (emission from an initial rather than final leg). The diagram $c)$ gives null contribution, because $\chi_3^\dagger \tau^- \chi_1 = 0$, being nucleons $3$ and $1$ both neutrons. 

Therefore the total amplitude for the t-channel reads
\begin{equation}
\mathcal{M}_t = \mathcal{M}_a + \mathcal{M}_b = \frac{iGg \, g_A}{\sqrt{2}} \frac{\ell^i}{\omega} \chi_3^\dagger \sigma^j \chi_1 \chi_4^\dagger \Big(\sigma^j \sigma^i - \sigma^i\sigma^j\Big)\chi_2 = \frac{\sqrt{2} Gg \, g_A l^i}{\omega}\epsilon^{ijk}\chi_3^\dagger \sigma^j \chi_1 \chi_4^\dagger \sigma^k \chi_2,
\end{equation}
where in the last step we used the relation $\sigma^j \sigma^i - \sigma^i\sigma^j = -2 i \, \epsilon^{ijk} \, \sigma^k$ and where we notice that the vector part has canceled out. In the non-relativistic theory is then very easy to compute the amplitude squared summed over spins, which in this particular case reads
\begin{equation}\label{Eq:TchannelLandau}
    \sum_{\rm spins} |\mathcal{M}_t|^2 = \frac{16 G^2 g^2 g_A^2}{\omega^2}\omega_1\omega_2 \epsilon^{ijk} \epsilon^{ij'k'} \underbrace{\rm Tr( \sigma^j \sigma^{j'}) \rm Tr(\sigma^k \sigma^{k'})}_{4 \, \delta^{jj'}\delta^{kk'}} = 384 \frac{G^2g^2g_A^2}{\omega^2}\omega_1\omega_2,
\end{equation}
where we used $\sum_{\rm spins} l^il^j = 8 \omega_1 \omega_2 \delta^{ij}$ for the leptonic current and in the last step we also used the relation of the Levi-Civita tensor $\epsilon^{ijk} \epsilon^{ijk} = 6$. 

We observe that the contribution from the u-channel, corresponding to the diagrams $d)$ to $f)$ in Fig.~\ref{DiagramMURCA}, mirrors that of the t-channel, while the interference between the two channels cancels out when summing over spins, as terms with an odd number of Pauli matrices always vanish. Consequently, the final result for the amplitude squared of these Landau parameters is simply twice the value given in Eq.~\ref{Eq:TchannelLandau}
\begin{equation}
    \sum_{\rm spins} |\mathcal{M}|^2 = 
768 \frac{G^2g^2g_A^2}{\omega^2}\omega_1\omega_2,
\end{equation}
in agreement with Eq.~39 of \citetalias{Friman:1979ecl}.

We now pass to consider the OPE term in the nucleon-nucleon potential
\begin{equation*}\label{Eq:OPEPotential}
V(\mathbf{k})^{\rm OPE} =  h'_{\pi,k}  (\boldsymbol{\sigma}_1 \cdot \boldsymbol{\hat{\bk}}) (\boldsymbol{\sigma}_2 \cdot \boldsymbol{\hat{\bk}}) \boldsymbol{\tau}_1 \cdot \boldsymbol{\tau}_2,
\end{equation*}
where
\begin{equation}
h'_{\pi,k} \equiv - \frac{f_\pi^2}{m_\pi^2} \frac{k^2}{k^2 +m_\pi^2}.
\end{equation}
In this case the amplitudes for the t-channel diagrams $a)$ and $b)$ with neutral pion exchange read
\begin{eqnarray}
\mathcal{M}_a = \frac{iG \ell^\mu}{\sqrt{2} \omega} \chi_3^\dagger h'_{\pi,k} (\bm{\sigma} \cdot \hat{\bk}) \tau^a \chi_1 \chi_4^\dagger \tau^- \left[ \delta_{\mu 0} - g_A \delta_{\mu i}\sigma^i\right] \tau^a (\bm{\sigma} \cdot \hat{\bk})  \chi_2, \\
\mathcal{M}_b = \frac{iG}{\sqrt{2}} \left( -\frac{\ell^\mu}{\omega} \right) \chi_3^\dagger h'_{\pi,k} (\bm{\sigma} \cdot \hat{\bk}) \tau^a \chi_1 \chi_4^\dagger \tau^a (\bm{\sigma} \cdot \hat{\bk}) \tau^- \left[ \delta_{\mu 0} - g_A \delta_{\mu i} \sigma_i \right] \chi_2.
\end{eqnarray}
One can verify that the isospin structure forces $a =3$ and that $\chi_4^\dagger \tau^- \tau^3 \chi_2 =1$, while $\chi_4^\dagger \tau^3 \tau^- \chi_2 = - 1$. Therefore we have
\begin{eqnarray}
\mathcal{M}_a + \mathcal{M}_b &=& \frac{\sqrt{2} \, i G l^\mu}{\omega}h'_{\pi,k} \chi_3^\dagger (\bm{\sigma} \cdot \hat{\bk})\chi_1 \chi_4^\dagger \left[ \delta_{\mu 0}(\bm{\sigma} \cdot \hat{\bk}) - \frac{1}{2}g_A \delta_{\mu i}  \left\lbrace \sigma^i, \sigma^j \right\rbrace \hat{\bk}^j\right]\chi_2  \\ \nonumber
&=& \frac{\sqrt{2} \, i G l^\mu}{\omega}h'_{\pi,k} \chi_3^\dagger (\bm{\sigma} \cdot \hat{\bk})\chi_1 \chi_4^\dagger \left[ \delta_{\mu 0}(\bm{\sigma} \cdot \hat{\bk}) - g_A \delta_{\mu j} \hat{\bk}^j\right]\chi_2.
\end{eqnarray}

For the diagram $c)$ we have instead
\begin{equation}
\mathcal{M}_c = \frac{iG}{\sqrt{2}} \left( -\frac{\ell^\mu}{\omega} \right) h'_{\pi,k}\chi_4^\dagger \tau^a (\bm{\sigma} \cdot \hat{\bk}) \chi_2 \chi_3^\dagger \tau^a (\bm{\sigma} \cdot \hat{\bk}) \tau^- \left[\delta_{\mu0} - \delta_{\mu i} \sigma_i\right] \chi_1,
\end{equation}
and one can check that in this case the isospin structure is such that $\tau^1$ and $\tau^2$ contribute both in the same way. Therefore we are left with 
\begin{equation}
\mathcal{M}_c = -\frac{\sqrt{2} \, i G l^\mu}{\omega}h'_{\pi,k}\chi_4^\dagger (\bm{\sigma} \cdot \hat{\bk}) \chi_2 \chi_3^\dagger (\bm{\sigma} \cdot \hat{\bk}) \left[\delta_{\mu0} - \delta_{\mu i} \sigma_i\right] \chi_1.
\end{equation}
Summing everything we notice that the vector part cancels and the final amplitude is
\begin{eqnarray}
\mathcal{M}^{\rm OPE}_t &=& \frac{\sqrt{2} \, i G \, g_A l^i}{\omega}\Big(\frac{f_\pi}{m_\pi}\Big)^2\frac{k^2}{k^2+m_\pi^2} \left[\hat{\bk}_i\chi_3^\dagger (\bm{\sigma} \cdot \hat{\bk})\chi_1 \chi_4^\dagger\chi_2 - \chi_4^\dagger (\bm{\sigma} \cdot \hat{\bk}) \chi_2 \chi_3^\dagger (\bm{\sigma} \cdot \hat{\bk})\sigma_i \chi_1\right]= \\ &=& \nonumber \frac{\sqrt{2} i G \, g_A l^i}{ \omega}\Big(\frac{f_\pi}{m_\pi}\Big)^2\frac{k^2}{k^2+m_\pi^2} \left[\hat{\bk}_i\chi_3^\dagger (\bm{\sigma} \cdot \hat{\bk})\chi_1 \chi_4^\dagger\chi_2 - \chi_4^\dagger (\bm{\sigma} \cdot \hat{\bk}) \chi_2 \chi_3^\dagger\chi_1 \hat{\bk}_i \right.\\
 &~&+ \left. i \epsilon^{ijk} \chi_4^\dagger (\bm{\sigma} \cdot \hat{\bk}) \chi_2 \chi_3^\dagger\sigma_k\chi_1 \hat{\bk}_j\right],
\end{eqnarray}
where in the second line we used the relation $\sigma_j\sigma_i = \delta_{ji} - i \epsilon^{ijk} \sigma_k$. This amplitude coincides with Eq.~32 of \citetalias{Friman:1979ecl}. Squaring and summing over spins we get
\begin{eqnarray}
\sum_{\rm spins} |\mathcal{M}^{\rm OPE}_t |^2 &=& \frac{16 \, G^2g_A^2 \, \omega_1\omega_2}{\omega^2}\Big(\frac{f_\pi}{m_\pi}\Big)^4 \Big(\frac{k^2}{k^2+m_\pi^2}\Big)^2\left[4\, \mathrm{Tr}(\sigma^i\sigma^{i'}) \, \hat{\bk}_i\hat{\bk}_{i'} +\right.\\
&~&+\left.\epsilon^{ijk}\epsilon^{ij'k'} \mathrm{Tr}(\sigma^a\sigma^{a'})\mathrm{Tr}(\sigma^j\sigma^{j'})\hat{\bk}_a\hat{\bk}_{a'}\hat{\bk}_j\hat{\bk}_{j'}\right] \nonumber \\  &=& \frac{256 \, G^2g_A^2 \, \omega_1\omega_2}{\omega^2}\Big(\frac{f_\pi}{m_\pi}\Big)^4\Big(\frac{k^2}{k^2+m_\pi^2}\Big)^2,
\end{eqnarray}
which is half of the result in Eq.~39 of \citetalias{Friman:1979ecl}. The u-channel is the same with $l$ instead of $k$ 
\begin{equation}
\sum_{\rm spins} |\mathcal{M}^{\rm OPE}_u |^2 = \frac{256 \, G^2g_A^2 \, \omega_1\omega_2}{\omega^2}\Big(\frac{f_\pi}{m_\pi}\Big)^4\Big(\frac{l^2}{l^2+m_\pi^2}\Big)^2;
\end{equation}
in triangular approximation $k \sim l \sim p_n$, so that the two contributions are the same. This explains the factor of 2 in Eq.~39 of \citetalias{Friman:1979ecl}, which therefore takes into account both t and u channels, neglecting the interference between the two channels. The latter can be easily computed as follows
\begin{equation}
\begin{split}
    2 &\sum_{\rm spins} \mathcal{M}^{\rm OPE}_t \mathcal{M}^{\rm OPE, \dagger}_u = -\frac{32 \, G^2g_A^2 \, \omega_1\omega_2}{\omega^2}h'_{\pi,k}h'_{\pi,l}\Big(\hat{\bk}_i\chi_3^\dagger (\bm{\sigma} \cdot \hat{\bk})\chi_1 \chi_4^\dagger\chi_2 - \chi_4^\dagger (\bm{\sigma} \cdot \hat{\bk}) \chi_2 \chi_3^\dagger\chi_1 \hat{\bk}_i  + \\ 
    &+i \epsilon^{ijk} \chi_4^\dagger (\bm{\sigma} \cdot \hat{\bk}) \chi_2 \chi_3^\dagger\sigma_k\chi_1 \hat{\bk}_j\Big)\Big(\hat{\bl}_i\chi_3^\dagger (\bm{\sigma} \cdot \hat{\bl})\chi_2 \chi_4^\dagger\chi_1 - \chi_4^\dagger (\bm{\sigma} \cdot \hat{\bl}) \chi_1 \chi_3^\dagger\chi_2 \hat{\bl}_i  + i \epsilon^{ijk} \chi_4^\dagger (\bm{\sigma} \cdot \hat{\bl}) \chi_1 \chi_3^\dagger\sigma_k\chi_2 \hat{\bl}_j\Big)^\dagger \\ 
    &=  -\frac{32 \, G^2g_A^2 \, \omega_1\omega_2}{\omega^2}h'_{\pi,k}h'_{\pi,l}\Big(\underbrace{\epsilon^{ijk}\epsilon^{ij'k'}}_{\delta^{jj'}\delta^{kk'}-\delta^{jk'}\delta^{j'k}}\text{Tr} \left[ (\bm{\sigma} \cdot \hat{\bk}) \sigma_{k'} \sigma_k (\bm{\sigma} \cdot \hat{\bl}) \right]\hat{\bk}_j \hat{\bl}_{j'} +\\
    &\qquad\qquad\qquad\qquad - 2\,i \hat{\bk}_i \hat{\bl}_{j'}\epsilon^{ij'k'} \text{Tr} \left[ (\bm{\sigma} \cdot \hat{\bk})(\bm{\sigma} \cdot \hat{\bl})\sigma_{k'}\right] + 2\,i \hat{\bk}_i \hat{\bl}_{j'}\epsilon^{ij'k'} \text{Tr} \left[ (\bm{\sigma} \cdot \hat{\bk})\sigma_{k'}(\bm{\sigma} \cdot \hat{\bl})\right] \Big) \\
    &= \frac{64 \, G^2g_A^2 \, \omega_1\omega_2}{\omega^2}\Big(\frac{f_\pi}{m_\pi}\Big)^4\Big(\frac{k^2}{k^2+m_\pi^2}\Big)\Big(\frac{l^2}{l^2+m_\pi^2}\Big)\left[-3 + (\hat{\bk}\cdot \hat{\bl})^2\right] \,\, \nonumber, \\
\end{split}
\end{equation}
which coincides with the third term in Eq.~71 of \citetalias{Friman:1979ecl} upon using the Lagrange's identity for their cross-product squared.

\bibliographystyle{bibi}
\bibliography{biblio}

\end{document}